\documentclass[pre,showpacs,preprint]{revtex4}
\usepackage{graphicx}
\usepackage{natbib}
\usepackage{amssymb}
\usepackage{amsmath}
\usepackage{verbatim}

\begin{document}

\title{Random, blocky and alternating  ordering in supramolecular polymers of chemically bidisperse  monomers }

\author{S. Jabbari-Farouji$^{1,2,3}$ and Paul van der Schoot$^{1,4}$  }

\affiliation{$^{1}$Theory of Polymer and Soft Matter Group, Eindhoven University of Technology, P.O. Box 513,5600 MB Eindhoven, The Netherlands }
\affiliation{$^{2}$ Dutch Polymer Institute, P.O. Box 902, 5600 AX Eindhoven, The Netherlands}
\affiliation{$^{3}$ LPTMS, CNRS and Universit$\acute{e}$ Paris-Sud, UMR8626, Bat. 100, 91405 Orsay, France}
\affiliation{$^{4}$Institute for Theoretical Physics, Leuvenlaan 4, 3584 CE Utrecht, The Netherlands}
\date{\today}

\begin{abstract}
As a first step to understanding the role of molecular or chemical polydispersity in self-assembly, we put forward a coarse-grained model that describes the spontaneous formation of quasi-linear polymers in solutions containing two self-assembling species.  Our theoretical framework is based on a two-component self-assembled Ising model in which the bidispersity  is parameterized in terms of the strengths of the binding free energies that depend on the monomer species involved in the pairing interaction. Depending upon the relative values of the binding free energies involved, different morphologies of assemblies that include both components are formed, exhibiting paramagnetic- , ferromagnetic- or anti ferromagnetic-like order,i.e., random, blocky or alternating  ordering of the two components in the assemblies. Analyzing the model for the case of ferromagnetic ordering, which is of most practical interest, we find that the transition from conditions of minimal assembly to those characterized by strong polymerization can be described by a critical concentration that depends on the concentration ratio of the two species. Interestingly, the  distribution of monomers in the assemblies is different from that in the original distribution, i.e., the ratio of the concentrations of the two components put into the system. The monomers with a smaller binding free energy are more abundant in short assemblies and monomers with a larger binding affinity are more abundant in longer assemblies. Under certain conditions the two components congregate into separate supramolecular polymeric species and in that sense phase separate. We find strong deviations from the expected growth law for supramolecular polymers even for modest amounts of a second component, provided it is chemically sufficiently distinct from the main one.

\end{abstract}

\maketitle

\newpage

\section{Introduction}

Self-assembly processes play a key role in the construction of biological structures and materials including the cell skeleton, viruses, bone, protein complexes and amyloid fibrils \cite{Nico1,Nico2,McPherson,Lehn,de-Greef,Koopmans}. These natural supramolecular structures have inspired scientists to exploit  supramolecular self-assembly principles as a tool in order to design and build bio-mimetic molecular structures from synthetic molecular compounds for purposes such as drug delivery and biomedical diagnostic technologies \cite{Ciferri, Weiss,Koop}. These novel materials have found applications in nanotechnology, medicine, including dental, cosmetic surgery and orthopedic applications, and personal care products \cite{SupraSA1,DNA-origami,Amalia-PNAS, EP1,EP2,EP3,EP4}.

Application of bio-inspired materials requires mass production of their molecular building blocks  with  efficient methods, which are not necessarily as accurate as the ones employed in research laboratories, or, for that matter, biology. Indeed, industrially produced self-assembling molecular units tend not to be very monodisperse, i.e., consist only of a single compound, but often consist of a large number of similar molecules with varying size, charge, chemical composition, and so on. Molecules not chemically identical to the target molecule are sometimes called impurities, but the whole collection may also be seen in some generalized sense as a polydisperse one \cite{polydisperse,polydisperse1}.

Taking as an example beta-sheet forming peptides, polydispersity of this kind may lead to inter-molecular binding affinities that vary as a function of the polydispersity attribute of interest. This may in this case include amino acid sequence, the number of amino acids that make up a peptide, small molecule reaction products that are able to hydrogen bond to the peptides and so on \cite{Amalia-PNAS}. All of this may have a large impact on the solution properties of the assemblies and on the structure of the assemblies themselves. Therefore, understanding the role of polydispersity on the nature of order-disorder transitions and morphologies of spontaneously formed supramolecular structures is very important  from both a fundamental and an applied scientific point of view.

Despite the fact that self-assembling molecular blocks are always to some degree chemically polydisperse, this issue has, as far as we are aware, received little attention in the literature. The influence of polydispersity has been studied theoretically  in the context of self-assembling block copolymers \cite{diblock1,diblock2,diblock3,diblock4}, and theoretically and experimentally in that of chiral amplification in supramolecular polymers \cite{Gestel,Markvoort,Smulders2,DNA1,DNA2,Yashin,Litmanovich1,Litmanovich2}. In the latter, the net observed chirality of a solution is studied by varying the composition in mixtures of enantiomers of self-assembling compounds and in mixtures of achiral and chiral species that co-assemble \cite{Markvoort,Smulders1}. Because studies like these focus entirely on the net macroscopic helicity, very little information is available on the structure and composition of the assemblies.

The aim of the present work is to get insight into the role of polydispersity in  the simplest case of self-assembly, i.e., that of quasi-linear self-assemblies  also called equilibrium polymers or EPs. In EPs all the monomers are able to bond  reversibly with each other with a binding free energy and form quasi-linear self-assemblies of varying length. The growth of these assemblies usually takes place in the form of nucleated assembly in which, due to a required change of conformation of monomers in the bound state, an activation free energy barrier must be overcome in order to form assemblies. As one increases the density of free monomers in the system, one observes a transition from a regime of mainly monomeric state to a  self-assembly dominated regime. The transition from one regime to another can be characterized by a critical concentration that depends on both nucleation and binding  free energies \cite{nyrkova,SupraSA2,Paul}.

An example of such systems is given by beta-sheet-forming peptides in which the binding between the peptides takes place through hydrogen bonding between the oxygen in the backbone of  the amino acid residues in one peptide and nitrogen in the backbone of  amino acids in the other peptide \cite{Amalia-PNAS,Koopmans}. The activation free energy in this case originates from the transformation of conformation of peptides from a random coil or alpha-helical state with a lower free energy to a an extended, rod-like conformation with a higher free energy. Nevertheless, peptides joined to a beta-sheet gain a binding free energy that is large enough to counterbalance the free energy of the rod-like bound state.

As a first step towards understanding the influence of polydispersity in linear self-assemblies, we focus our theoretical study on  a \textit{bidisperse} self-assembling system, schematically illustrated in Fig. 1. To model linear self-assembly in such systems, we propose a two-component lattice-gas model in which bidispersity is incorporated through the species-dependent free energy parameters describing the binding and nucleation processes \cite{Gestel}. Mapping the problem onto a 1-D Ising model and invoking the standard transfer matrix method, we calculate the partition function and explore the ordering of the two components in assemblies of arbitrary length. This allows us to probe the composition assemblies of all sizes as function of concentration and interaction strengths between bound monomers.

We note that, in spirit at least, our approach is similar to that found in other works in the wider context of polymer physics, including the helix-coil transition of polypeptides and the melting transition of DNA \cite{Zimm-Bragg,DNA1,DNA2}, as well as structural transformations in two-component copolymers \cite{Yashin}, where transfer matrix methods have been employed. Furthermore, lateral ordering in tape-like structures of annealed blocky copolymers has been studied by simulations \cite{Litmanovich1,Litmanovich2}. The difference with the earlier work lies in the coupling between self-assembly, composition and length distribution of the assemblies.

The binding free energies describing our equilibrium polymerization model are denoted $b_{ij}>0 $, measuring the free energy \textit{gain} (in units of thermal energy) of the bonded interaction between two monomers of type $i,j=1,2$, hence the positive sign, where $b_{12} = b_{21}$ by symmetry. For definiteness, we presume that $b_{11}>b_{22}$, so molecules of species 1 forge stronger bonds to each other than species 2 do. There are also two activation free energies $a_i$ associated with the two species $i=1,2$ describing the (dimensionless) free energy penalty associated with conformational changes when monomers are absorbed in assemblies. See Fig. 1.

Our aim is to answer the following two three central questions:
\begin{itemize}
\item[i)] What happens to the mean size of the assemblies if we mix two distinct self-assembling species that are able to co-assemble?
\item[ii)] Do the two components actually co-assemble into linear aggregates, or  do they form  chemically pure linear assemblies consisting of one species only?
\item[iii)] What physical principles regulate the composition of the assemblies?
\end{itemize}
We find that the relevant quantity determining the arrangement of monomer species in the assemblies (the morphology) is an effective ``coupling constant'' $J \equiv\frac{1}{4}(b_{11}+b_{22}-2 b_{12})$ in the parlance of the Ising  model, which  depends on a linear combination of the binding free energies $b_{ij}$. This coupling constant is, of course, not unlike the Flory-Huggins parameter in binary lattice fluids \cite{Rubinstein}
Depending on the value of this effective coupling constant $J$, three different morphologies can be envisaged. For $J > 0$, the two types of monomer are organized in linear self-assemblies in a blocky, ``ferromagnetic-like'' order. The ``paramagnetic'' $J = 0$ case corresponds to a random distribution  of monomers in  the assemblies, whereas $J < 0$  leads to ``anti-ferromagnetic-like'', alternating ordering of the monomers along the assemblies. In the limit $J \ll - 1$, the latter would represent co-ordination polymers \cite{Vermonden}.
\begin{figure}[h!]
\begin{center}
\includegraphics[scale=0.4]{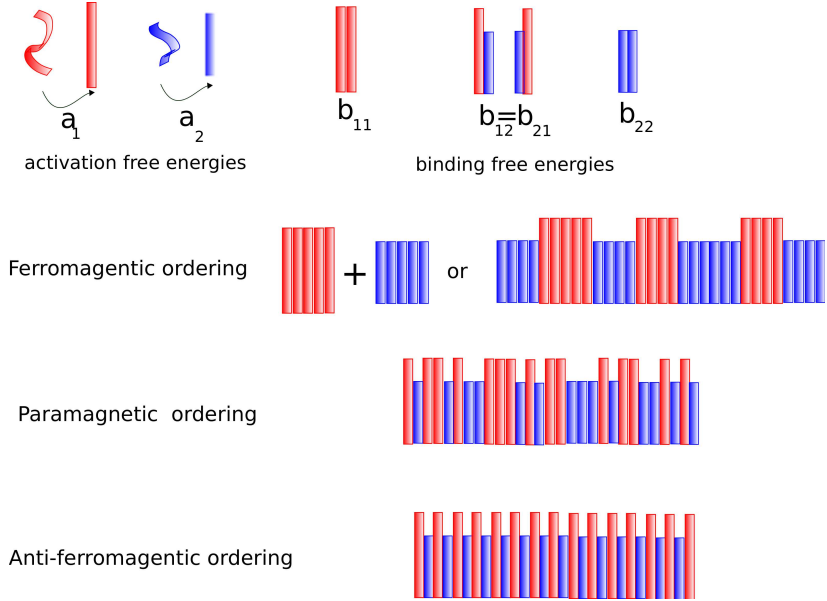}
\caption{A schematic of the linearly self-assembling system under study and the morphology of the structures formed. The two species of self assembler give rise to two activated states with associated free energies $a_1$ and $a_2$, accounting for conformational changes necessary for binding, and three binding free energy gains $b_{11}$, $b_{12}=b_{21}$ and $b_{22}$
\label{fig1}}
\end{center}
\end{figure}
The ``paramagnetic'' (random) case corresponds to a self-averaging system that exactly behaves like a monodisperse system and therefore is not of interest in our work.
The ``ferromagnetic'' (blocky) case occurs when the binding-free energy between the two distinct species is less than the average binding free energy of the two species. This is the case that is relevant to those situations we are interested in, i.e., those cases where polydispersity as defined earlier plays a role. For instance, for the beta-sheet forming oligopeptides discussed earlier, and reported on by Aggeli and collaborators, binding takes place through hydrogen bonding \cite{Amalia-PNAS}. When a short peptide binds to a longer one, the number of hydrogen bonds is equal to that formed between two short ones, suggesting that $b_{12}=b_{22}$ and $J = (b_{11}-b_{22})/4>0$, so this would indeed correspond to the ``ferromagnetic'' (blocky copolymeric) case.

Although the ``ferromagnetic'' case $J>0$ does obviously not cover all possible circumstances, one can envisage many experimental situations where this condition is met. For completeness we present our results for arbitrary values of $J$ throughout the paper although we do analyze our model and discuss the consequences of bidispersity only for the ferromagnetic case. According to our findings, the bidisperse system  behaves in many ways similar to the monodisperse one, and exhibits  a transition from a regime with minimal assembly to one where self-assembly predominates. If sufficiently co-operative, the polymerization transition is sharp and is demarcated by a critical concentration or temperature. The precise value of the critical concentration or temperature depends on the stoichiometric ratio of the concentrations of the two components involved.

Furthermore, for large asymmetries between the two species, i.e., large $J$ values, and for sufficiently low concentrations, a ``demixing'' region appears where pure self-assemblies, composed mainly of one species, coexist. Note that there is no demixing here on macroscopic scale, only on the scale of the individual assemblies. Considering the distribution of the two species along the assemblies, we find that the distribution of the two species in the assemblies differs from that of the parent distribution as the size of the assemblies grows. In short assemblies the population of species with less binding affinity is dominant, while in longer assemblies monomers with a bigger binding affinity are more abundant.

The remainder of this paper is organized as follows. In Section 2, we first outline a theoretical framework that is based on a two-component self-assembled lattice-gas model. In section 3, we show that mapping our model onto an Ising model allows us to  explore and predict the ``phase'' behavior  of bidisperse  monomers forming quasi-linear self-assemblies.  In Section 4, we explore the self-assembly behavior in the low- and high-concentration limits. Sections 5 and 6 are devoted to  a comprehensive analysis of our model, focusing in particular on a bidisperse system with ferromagnetic-like (blocky copolymeric) ordering. We discuss and compare the results for the cases of small and large $J$. Of particular interest is the relative fraction of the two species in the assemblies as their size varies.  Finally, we conclude our work in section 7, where we summarize our main findings and discuss and compare the influence of bidispersity in self-assembling systems with  other thermodynamical systems.

\section{Equilibrium statistics of bidisperse self-assembling monomers}
We consider a model system consisting of two self-assembling species that can form linear self-assemblies of any length $N = 1,2, ..., \infty$ . Each monomer in a typical linear self-assembly can be either of species 1 or 2.  The difference between  the two species arise either from their size, chemical structure and so on. We parameterize this difference between the species  by invoking effective activation and binding free energies that depend on the type of species. Again, let $b_{ij}>0 $ denote the free energy gain of the bonded interaction between two monomers of type $i,j=1,2$ in a self-assembly, and  $a_{i}>0 $ the activation free energy of a monomer of type $i$. In principle, the value of the former depends on the next neighbors of the two binding molecules but we ignore this complication here.
In our model, $b_{ij}=b_{ji}$ and $a_i > 0$, where for definiteness we suppose $b_{11}>b_{22}$. All free energies are scaled to the thermal energy $k_BT \equiv 1$. Here, $k_B$ denotes the Boltzmann constant and $T$ the absolute temperature.

More generally, these effective free energies can originate from different types of interactions such as electrostatic, hydrogen bonding, hydrophobic and van der Waals interactions. For instance, for the pertinent system of Aggeli et al. mentioned earlier \cite{Amalia-PNAS}, involving peptides of $L$ amino-acids as monomers, the binding free energy gains $b_{ij}$ result from hydrogen bonds between the oxygen atoms in the backbone of one peptide and the nitrogen atoms in the  backbone of the other peptide. This gives rise to $2$ hydrogen bonds per residue and $2L$ hydrogen bonds in total, suggesting that $b_{ii}= b_0 L$, with $b_0$ a proportionality constant. Similarly, $b_{12}$ results from the hydrogen bonding between the backbones of two peptides of different length, and the number of hydrogen bonds in this case is equal to twice the number of residues of the shorter peptide, so $b_{12}=b_{22}$.

The transformation free energies $a_{i}$ result from loss of conformational entropy due to a change of configuration of peptides from a single, coil-like monomer to a rod-like extended and a bound state in the assembly. As a first order approximation, one would expect them to depend linearly on the size of oligopeptides, i.e., $a_i=a_{0} L_i$.

The (dimensionless) grand potential energy of a system consisting of free monomers and self-assembled polymers can be written as the sum of an ideal entropy of mixing and the contribution of the internal partition functions of all the assemblies of varying size $N$, giving
\begin{equation}
\label{eq:a0}
\frac{\Omega}{V}= \sum_{N=1}^{\infty} \rho(N)[\ln (\rho(N)\nu)-1 -\ln Z_N(\mu_i, b_{ij},a_{i})],
\end{equation}
where $V$ is the volume of the system, $Z_N$ the (semi-grand partition) function of an assembly consisting of $N$ monomers. Here, $\nu$ denotes the interaction volume equal roughly to an effective volume of a solvent molecule \cite{Paul}, which we assume to be independent of the type of species, and $\rho (N)$ the number density of self-assemblies of size $N$. The semi-grand partition function $Z_N$ counts the number of configurational states of linear assemblies of size $N$, composed of the two species of monomer and described by their relevant binding and transformation free energies, and dimensionless chemical potentials $\mu_i$ ($i=1,2$). The latter are fixed by the  total concentration of  the monomers in the solution. Our grand potential tacidly assumes the assemblies to be in the dilute limit, i.e., interactions between assemblies are presumed to be negligible.

The equilibrium size distribution minimizes the grand potential Eq. \ref{eq:a0}. Setting $\delta \Omega / \delta \rho (N)=0$ yields
\begin{equation}
\label{eq:distribtion}
 \rho(N)= \nu^{-1}Z_N(\mu_i, b_{ij}, a_{i}).
\end{equation}
Therefore, our task of finding the equilibrium size distribution reduces to the calculation of the semi-grand partition function of a polymer with $N$ degrees of polymerization, whose monomers can be either of the two species.  To do this, we model  a linear self-assembly of length $N$ as an one-dimensional interacting two-component lattice gas of size $N$ each site of which is occupied by either of species 1 or 2. Each site numbered $l=1,...,N$ can be identified by  the occupation numbers $n_l^{(1)}=0,1$ and $n_l^{(2)}=1-n_l^{(1)}$ describing the number of monomers of species 1 and 2 at position $l$ along the lattice, respectively. These occupation numbers automatically obey the constraint  $\Sigma_{l=1}^{N}( n_l^{(1)}+n_l^{(2)})=N$.

The mutual exclusivity of occupation of each site by either of the species gives us the possibility of mapping the two occupation numbers to a single ``spin'' variable of the  Ising model of ferromangetism by a simple transformation
\begin{eqnarray}
\label{eq:a1}
n_l^{(1)}=\frac{1}{2}(1+S_l),\\
n_l^{(2)}=\frac{1}{2}(1-S_l),
\end{eqnarray}
where $S_l$ are the ``spin'' variables that take $\pm1$ values. As a result, in this representation the  species 1 can be though of as an ``up'' spin and the species 2 as a ``down'' spin. In order to calculate the semi-grand partition function $Z_N$, we first express the Gibbs free energy of a self-assembly of length $N$ in terms of the spin variables. The (dimensionless) internal free energy of monomers of type $i$ in free solution is lower by an amount equal to $- a_{i}$ compared to those of bonded ones due to the larger number of available degrees of freedom in the unbound state.
\begin{eqnarray}
\label{eq:a2}
 G(1)=-\frac{a_{1}}{2}(1+S_1)-\frac{a_{2}}{2}(1-S_1)\\
G(N>1)=\sum_{i=1}^{N-1} [-\frac{b_{11}}{4}(1+S_i)(1+S_{i+1})-\frac{b_{22}}{4}(1-S_i)(1-S_{i+1})\\ \nonumber
-\frac{b_{12}}{4}\{(1+S_i)(1-S_{i+1})+(1-S_i)(1+S_{i+1})\}]
\end{eqnarray}
The semi-grand partition function can then be written as
\begin{equation}
\label{eq:a3}
Z_{N}(\mu_i, b_{ii},a_{i},b_{12})=\sum_{\{S_l\}} \exp [- G(N)+ \sum_{l=1}^{N}(\mu_1 \frac{(1+S_l)} {2}+\mu_2 \frac{(1-S_l)} {2})].
\end{equation}
The partition function for monomers in the unbound, solution state has the simple form of $Z_{1}=\exp(\mu_1+a_{1})+\exp(\mu_2+a_{2})$.

For all cases $N>1$, we find, by carrying out the sum over the spin values and simplifying the semi-grand partition function by collecting terms of equal order, that it is equivalent to the partition function of an Ising chain of size $N$. Its Hamiltonian reads
\begin{equation}
\label{eq:a5}
 \mathcal{H}_{N > 1}(\{S_l\})=-J\sum_{l=1}^{N-1} S_l S_{l+1}-H \sum_{l=1}^{N}S_l- \varepsilon_0(N)+\frac{1}{4}(b_{11}-b_{22})(S_1+S_N),
\end{equation}
with the effective coupling constant $J$, magnetic field strength $H$ and an energy term $\varepsilon_0(N)$ that is an invariant of the spin states, defined as
\begin{eqnarray}
\label{eq:a6}
J\equiv\frac{1}{4}(b_{11}+b_{22}-2b_{12}),\\
H\equiv \frac{1}{2}[(b_{11}-b_{22})+(\mu_1-\mu_2)], \\
\varepsilon_0(N)\equiv \frac{(N-1)}{4}(b_{11}+b_{22}+2b_{12})+ \frac{N}{2}(\mu_1+\mu_2),
\end{eqnarray}
In the special case of $b_{22}=b_{12}$, relevant for the oligopeptides of Aggeli et al. \cite{Amalia-PNAS} discussed earlier, we obtain  $J=\frac{1}{4}(b_{11}-b_{22})$ and $\varepsilon_0(N)= (N-1)(b_{11}+3b_{22})/4+ N(\mu_1+\mu_2)/2$.

If we introduce the renormalized chemical potentials $\mu'_i \equiv \mu_i + b_{ii}$, this gives $H=(\mu'_1-\mu'_2)$ and $\epsilon_0(N)=N(-J+(\mu'_1+\mu'_2)/2)-\bar{b}$, where $\bar{b}=(b_{11}+b_{22}+2 b_{12})/4$ is as before the mean binding free energy. We conclude that there are only two relevant energetic parameters, $J$ and $\bar{b}$. The average binding free energy does \textit{not} couple to the spin states, is independent of $N$, and plays a role similar to that of the binding free energy in the monodisperse case. The parameter $J$ does couple to the spin states of the lattice and determines in the end the composition of an arbitrary assembly. In the next section, we describe how  to calculate the partition function of our model and other  relevant quantities such as  fraction of self-assemblies by  exploiting the  mapping onto the Ising model.

\section{Mapping onto the Ising model}
Mapping our problem onto the one-dimensional Ising model already provides us with a direct insight into the general ``phase behavior'' of the system at hand -- here ``phase behavior'' does not refer to phase separation on the macroscopic scale but rather on the microscopic scale, that is, between assemblies. As already advertised in section I, depending on the value of the coupling constant $J$, associated with the spin (or occupation) states of two neighboring sites, different types of ordering may appear in the assemblies. For $J>0$ ferromagnetic ordering is favored, implying blocky copolymers, for $J=0$ the paramagnetic case is favored, meaning random copolymers, and for $J <0$ anti-ferromagnetic ordering, associated with alternating, copolymeric ordering. See  Fig. \ref{fig1}.

Exploiting the standard method of the transfer matrix \cite{goldenfeld}, the resulting partition function for free boundary conditions (implying no preference for any monomer to sit at the ends of the assemblies) takes the form,
 \begin{eqnarray}
\label{eq:a7}
Z_{N>1}&=& [x_+ \lambda_+^{N-1}+x_{-}\lambda_-^{N-1}] \exp(\varepsilon_0(N)),  \\
\lambda_{\pm}&=&\frac{(1+e^{2H})e^{2J}\pm \sqrt{4 e^{2H}+(e^{2H}-1)^2 e^{4J}}}{2e^{H+J}},\\
x_{\pm}&=&  \frac{(-e^{H} + e^{2J} (z_1/z_2)^{1/2} - e^{H+J} \sqrt{z_1/z_2} \lambda_{\pm})  (e^{2J} + e^{H} (z_1/z_2)^{1/2} -
   e^{H+J} \lambda_{\mp})}{e^{2H+J}  (z_1/z_2)^{1/2}  (\lambda_{\mp} -\lambda_{\pm})},
\end{eqnarray}
in which  $z_i$  are defined as the fugacity of species $i$, that is, $z_i\equiv\exp(\mu_i)$, and $\lambda_{\pm}$ the eigenvalues of the transfer matrix. Note that our choice of (free) boundary conditions is expressed by the last term of Eq. \ref{eq:a5}, through the values of  $x_{\pm}$.

Our next step is to determine the size distribution of linear chains of assemblies $\rho(N)$ from Eq. \ref{eq:distribtion} in terms of the concentrations of the two species involved. To do so, we need to eliminate the chemical potentials $\mu_1$ and $\mu_2$ from Eq. \ref{eq:distribtion} and Eq. \ref{eq:a7}. The values of the chemical potentials can be established from  the conservation of mass for either of the two species,
 \begin{eqnarray}
\label{Mconservation}
\sum_{N=1}^{\infty}N \rho(N)\nu=\Phi_{1}+\Phi_{2}\equiv \Phi, \\
\sum_{N=1}^{\infty}N \rho(N)\nu \langle S(N)\rangle=\Phi_{1}-\Phi_{2},
\end{eqnarray}
where $\Phi_{1}$ and $\Phi_{2}$ are the molar fractions of species 1 and 2, respectively, and  $\langle S \rangle$ is the average spin value, which can be calculated  from  the  partition function as $\langle S(N) \rangle = (1/N) \partial \ln Z_N/\partial H $. Here, we have for convenience introduced the overall \textit{molar fraction} of both species $\Phi$. Note that the  low density approximation used in  Eq. \ref{eq:a0}implies that our results are only valid as long as  $\Phi_{i} \ll 1 $.

To calculate the above sums, we rephrase the partition function $Z_N(\mu_i, b_{ij},a_{i})$ in terms of intensive and extensive parts for the case $N >1$.  Eq. \ref{eq:distribtion} can be rewritten as,
\begin{eqnarray}
\label{eq:dist}
 \rho(1)\nu&=& z_1 \exp(a_{1})+ z_2 \exp(a_{2}), \\
 \rho(N>1)\nu &=& Z_N(z_i, b_{ij},a_{i})= \sum_{i=+,-} x_i e^{-\bar{b}}   \Lambda_i^N/ \lambda_i
\end{eqnarray}
where  $\Lambda_i \equiv \lambda_i \exp(\bar{b})\sqrt{z_1 z_2}$ with $\bar{b}=(b_{11}+b_{22}+2 b_{12})/4$.
Obtaining a convergent sum requires that $\Lambda_i <1$, which restricts the possible values of the chemical potentials. The two sums in Eq. \ref{Mconservation}  can now be calculated as a geometrical series and be simplified to
\begin{equation}
\label{eq:sum1}
  \Phi=z_1 \exp(a_{1})+ z_2 \exp(a_{2})+\sum_{i=+,-}x_i e^{-\bar{b}}\Lambda_i^2 \frac{2-\Lambda_i}{\lambda_i (1-\Lambda_i)^2}
\end{equation}
\begin{equation}
\label{eq:sum2}
 \Phi_{1}-\Phi_{2}=z_1 \exp(a_{1})- z_2 \exp(a_{2})
 + \sum_{i=+,-}x_i e^{-\bar{b}}\frac{\Lambda_i}{\lambda_i (1-\Lambda_i)^2}[-\Lambda_i(1-\Lambda_i)\frac{\partial\ln(x_i/\lambda_i)}{\partial H}+(2-\Lambda_i) \frac{\partial \Lambda_i} {\partial H} ].
\end{equation}
Equations \ref{eq:sum1} and \ref{eq:sum2} should be solved simultaneously to determine $z_i$ and the relevant quantities of interest such as the fraction of self-assemblies, the average degree of polymerization and so on. Therefore, these equations are the central equations that are used throughout the rest of this paper, when analyzing the model for different free energy parameters and concentrations of species.

To find the simultaneous solution of these two equations for known concentrations of both species, we need to look for solutions that satisfy the conditions $0< z_i < z^*_i <1$, where the $z^*_i$ are the maximum value for the fugacity of species $i$ to be discussed in the next section. However, to explore the dependence of phase behavior on the concentration of the two species,  our strategy is to vary  $0< z_i< z^*_i$  and calculate the corresponding molar fractions.
Once the chemical potentials from Eqs. \ref{eq:sum1}  and \ref{eq:sum2} are determined, it is straightforward to calculate the relevant quantities of interest, such as the number-averaged size of assemblies defined as,
 \begin{equation}
\label{eq:a8}
 \langle N \rangle =\frac{\sum_{N=1}^{\infty}N \rho(N)}{\sum_{N=1}^{\infty}\rho(N)}=\frac{\Phi}{\sum_{N=1}^{\infty}\rho(N) \nu},
\end{equation}
where the sum in the denominator can be easily calculated as a geometrical series,
 \begin{equation}
\label{eq:sum3}
 \sum_{N=1}^{\infty}\rho(N)\nu=z_1 \exp(a_{1})+ z_2 \exp(a_{2})+\sum_{i=+,-}x_i e^{-\bar{b}}\frac{\Lambda_i^2}{\lambda_i(1-\Lambda_i)}
\end{equation}
 Likewise, one can obtain the fraction of self-aggregates $f$,  defined as
\begin{equation}
\label{eq:a9}
 f=1-\frac{\rho(1)\nu }{\Phi}=1-\frac{z_1 \exp(a_{1})+ z_2 \exp(a_{2})}{\Phi}.
\end{equation}

Having Set up our model for the linear self-assembly of bidisperse monomers, we have  all the tools in hand for a calculation of the quantities of interest. To gain insight into the phase behavior of the system, we first look at a few limiting cases,  where analytical progress is possible. Therefore, the next section is devoted to the investigation of the self-assembly behavior in the  low and high concentration limits, and  a thorough  numerical investigation of our model is postponed to section after that, i.e., in section V.

\section{Limiting cases}

\subsection{The monodisperse case}
It is instructive to explore the behavior of system in the limiting case where one can simplify the equations and obtain analytical results. The results obtained in these cases shed some light on the general ``phase'' behavior of the system at hand.  Before we go into details it is important to notice that we recover the results for the monodisperse  case \cite{nyrkova} from the present theory in the limit $z_2 \rightarrow 0$, i.e., in the limit of zero concentration of second species.  In this  limit, $H\rightarrow \infty$, $\lambda_+ \rightarrow e^{J+H}$ and $\lambda_- \rightarrow 0$ yielding $\Lambda_+\rightarrow  z_1 \exp(b_{11}) $. It is important to point out that the second eigenvalue of the transfer matrix becomes identically zero due to the vanishing concentration of the second species. This is different from the usual ground-state approximation valid in the limit of large N, where the contribution of  the second eigenvalue to the partition function becomes negligible relative to the first.

We find that the number density of monomers and assemblies are
\begin{eqnarray}
\rho(1)\nu=z_1 \exp(a_{1}), \\ \nonumber
 \rho(N>1)\nu=\exp (-b_{11}) {\Lambda_+}^N ,
\end{eqnarray}
where $\Lambda_+$  is the solution of  the cubic equation of the form $\Lambda_+ K_{a1}  +  \exp(b_{11}) \Lambda_+ ^{2 }(2-\Lambda_+ )/(1-\Lambda_+ )^2= \Phi$  with $K_{a1}\equiv \exp(-a_{1})$ the activation constant.  This is identical to what was already known for monodisperse systems, confirming the consistency of our model \cite{nyrkova,Paul}. The number-averaged length of assemblies for sufficiently high concentrations $\Phi_1 \gg \Phi_1^*$ can be written as
 \begin{equation}
\label{eq:ave0}
\langle N\rangle = \frac{2-\Lambda_+}{1-\Lambda_+},
\end{equation}
which can be expressed in terms of the molar fraction of monomers in  the simple form of
 \begin{equation}
\label{eq:ave01}
\langle N\rangle \simeq \sqrt{(\Phi_1/\Phi_1^*-1)/K_{a1}} .
\end{equation}
Here, $\Phi_1^* \equiv \exp(-b_{11}+a_1)$ is a critical concentration beyond which one can consider the system in the assembly-dominated regime \cite{nyrkova,Paul}. Provided that $a_{1}\gg 1$, this leads to a  very sharp cooperative polymerization transition demarcated by $\Phi_1^*$. Eq.(\ref{eq:ave01}), and produces the well-known \textit{square-root growth law} of self-assembled polymers \cite{SupraSA2}.

The fraction of materials in  assemblies in the low and high concentration regions in that case obey
 \begin{eqnarray}
\label{eq:f1}
f& \simeq& 2 K_{a1} \Phi_1/\Phi_1^* \quad \quad \verb"if" \quad \Phi_1 \ll \Phi_1^* \\
f& \simeq &1-\Phi_1^*/\Phi_1  \quad \quad \verb"if" \quad \Phi_1 \gg  \Phi_1^*.
\end{eqnarray}
This shows that for low concentrations the fraction of assemblies grows linearly with concentration, with a slope proportional to the activation constant.  If $a_1\gg 1$ then $K_{a1}\ll 1$  and $f \rightarrow 0$ for $\Phi < \Phi_*$. On the other hand, for high concentrations the fraction of self-assemblies differs from unity by the amount equal to the ratio of the critical concentration to the total concentration of monomers. We refer to a recent review by one of us for a more detailed discussion of activated equilibrium polymerization \cite{Paul}. Next we consider the limiting behavior of bidisperse systems.

\subsection{The bidisperse case for $\Phi \gg \Phi^*$}
If the total concentration of monomers is large enough so that $\langle N \rangle \gg 1$, which is the case for $\Phi > \Phi_* $ when the activation constants $a_i$ are sufficiently large, we can apply the ground-state approximation for the partition function of the Ising model and, therefore, ignore the smaller eigenvalue. As a result, the number average length of tapes takes the simple form of
\begin{equation}
\label{eq:ave1}
\langle N\rangle_n \simeq \frac{2-\Lambda_+}{1-\Lambda_+} .
\end{equation}
Furthermore, this leads to a number of simplifications and allows us to get an insight into the high-concentration behavior of self-assembly. Particularly, we can find  an analytical expression for the fraction of self-assemblies. In this limit, the concentration of monomers contributing to the assemblies becomes dominant relative to that of free monomers and $\Lambda_+=\lambda_+ \exp(\bar{b})\sqrt{z_1 z_2}$ approaches its limiting value, i.e., $\Lambda_+\rightarrow 1$. As before, $\bar{b}$ denotes the average of the binding free energies. Therefore, the concentration of free monomers reaches its critical value $z_1^*\exp(a_{1})+ z_2^*\exp(a_{2})\equiv \Phi^*$, which corresponds to the maximum molar fraction of free monomers. In some specific cases, this maximum molar fraction of free monomers characterizes the transition from minimal assembly to an assembly-dominated regime. Again, provided that $a_i \gg 1$, this leads to a sharp cooperative polymerization transition.

The critical concentration depends, in principle, on  the stoichiometric ratio of the two components, $\alpha=\Phi_1/\Phi_2$, in addition to the  various free energy parameters that describe the model.  Relatively straightforward algebra gives for the fraction of self-assemblies  a universal curve for high concentrations and of the simple form
\begin{equation}
\label{eq:ave}
 f \simeq 1-\frac{\Phi^*(\alpha)}{\Phi} \quad \quad \quad \quad  \verb"if" \quad \frac{\Phi}{\Phi^*(\alpha)} \gg 1,
\end{equation}
where $\Phi_*(\alpha)$ denotes the critical concentration that now depends explicitly on the stoichiometric ratio $\alpha$. We can obtain the molar fraction of free monomers for each of the species, hence, the critical concentration as a function of $\alpha$, by solving the two following equations simultaneously,
\begin{eqnarray}
\label{eq:C*}
 \Lambda_+ (z_1^*, z_2^*) &=&1, \\ \nonumber
 \frac{\partial \Lambda_+}{\partial H} |_{\Lambda_+=1}&=&\frac{\alpha-1}{\alpha+1},
\end{eqnarray}
where the second equation results from simplifying  $ \Phi_1(z_1^*, z_2^*)/ \Phi_2(z_1^*, z_2^*)= \alpha$ in  the ground-state approximation.

We were not able to find an exact analytical expression for the free monomer molar fractions and critical concentration for arbitrary values of $\alpha$. However, for the special case of $J \gg 1$, we did obtain asymptotic expressions for $z_1^*$ and  $z_2^*$. The case $J \gg 1$ occurs  either if there is a large asymmetry  between the two species, or if the two species do not have a large affinity to bond to each other. In this limit,  one can simplify the above equations to obtain an asymptotic  solution for the fugacities and therefore, the density of free monomers.
\begin{eqnarray}
\label{eq:C*03}
 \Phi^f_1\simeq    \Phi_1^* \frac{2 e^{4J}(1+ \alpha)-\alpha-\sqrt{\alpha^2+4 (1+\alpha) e^{4J}}}{2 (1+\alpha) e^{4J}}
\end{eqnarray}
and
\begin{eqnarray}
\label{eq:C*04}
 \Phi^f_2\simeq   \Phi_2^* \frac{ \left(2+2 e^{4J} (1+\alpha )-2 \sqrt{\alpha ^2+4 e^{4J} (1+\alpha )}+\alpha  \left(2+\alpha -\sqrt{\alpha ^2+4 e^{4J} (1+\alpha )}\right)\right)}{2 e^{4J} (1+\alpha )}
\end{eqnarray}
 Here, $\Phi^*_i \equiv  \exp(-b_{ii}+a_{i})$ are critical concentrations already defined for the individual species.

For an arbitrary value of $J>0$,  we can find expressions only in the limits of small and large $\alpha$, corresponding to the respective limits  $e^{-H} \ll 1$ and $e^{H} \gg 1$ in the Ising model.
The molar fraction of free monomers of either of species in the high concentration regime are,
\begin{eqnarray}
\label{eq:C*11}
 \Phi^f_1\simeq \alpha  \Phi_1^* e^{4J} \quad \quad  \Phi^f_2 \simeq (1- \alpha)  \Phi_2^*  \quad\quad  \alpha \ll e^{-4J}, \\
 \label{eq:C*12}
 \Phi^f_1 \simeq  \Phi_1^* (1-\frac{1}{\alpha})  \quad  \quad  \Phi^f_2 \simeq \frac{1}{\alpha} \Phi_2^* e^{4J} \quad\quad  \alpha \gg e^{4J},
\end{eqnarray}
which agrees with the small and large $\alpha$ limits of Eqs. \ref{eq:C*03}-\ref{eq:C*04}, as they should.

An interesting conclusion can be drawn if we compare the ratio of molar fraction of free monomers of the two species, i.e., $\alpha^f \equiv \Phi^f_1/\Phi^f_2$ with $\alpha$.
\begin{eqnarray}
\label{eq:l1}
  \alpha^f/ \alpha \simeq e^{4J} \Phi_1^*/\Phi_2^*   \quad\quad  \alpha \ll e^{-4J} \\
\label{eq:l2}
 \alpha^f/ \alpha \simeq e^{-4J} \Phi_1^*/\Phi_2^* \quad\quad  \alpha \gg e^{4J}
\end{eqnarray}
These results show that, for both small and large $\alpha$ values,  the ratio of molar fraction of free monomers is different from $\alpha$, pointing to fractionation effects even in the ``high-concentration'' regime, that is, concentration very much higher than the critical one. We will discuss this issue further in the next section where we investigate  numerically the full concentration behavior of self-assembly.

As a result, we obtain the following expressions for the critical concentration of the bidisperse system.
\begin{eqnarray}
\label{eq:C*1}
 \Phi^*& \simeq &\Phi_2^*+ \alpha ( \Phi_1^* e^{4J}-\Phi_2^*) \quad\quad  \alpha \ll e^{-4J} \\ \nonumber
 \Phi^*& \simeq &\Phi_1^*+\frac{1}{\alpha} (\Phi_2^*e^{4J}-\Phi_1^* )  \quad\quad  \alpha \gg e^{4J}
\end{eqnarray}
These functional forms clearly demonstrate that  the critical concentration not only depends on the critical concentration values of the two species, and their ratio $\alpha$, but is also strongly  influenced by the value of coupling constant $\exp(4J)$. Notice that in the limits $\alpha \rightarrow 0$  and   $\alpha \rightarrow \infty$,  we recover the critical concentrations of species 2 and 1, respectively, verifying the self-consistency of our calculation.

From the above discussion, we recognize  four different  regimes: $\alpha \ll e^{-4J}$ and $\alpha \gg e^{4J}$, and for both of these, the cases $e^{4J} \ll 1$ and $e^{4J}\gg 1$. If $\alpha \ll e^{-4J}$ and $ e^{4J}\ll \Phi_2^*/\Phi_1^*$, the critical concentration decreases linearly upon increase of $\alpha$ with a slope proportional to $\Phi_2^*$, i.e., $\Phi^* \approx \Phi_2^*- \alpha \Phi_2^*$. However, for $\alpha \ll e^{4J}$ and $e^{4J} \gg \Phi_2^*/\Phi_1^*$, the slope is determined by $e^{4J}$, i.e., $\Phi^* \approx \Phi_2^*+ \alpha  \Phi_1^* e^{4J}$. This leads to an initial increase of the critical concentration for sufficiently small values of $\alpha$.

In the other extreme of $\alpha \gg e^{4J}$, we can also make a distinction between the cases $e^{4J} \ll \Phi_1^*/\Phi_2^*$ and $e^{4J} \gg \Phi_1^*/\Phi_2^*$. In the case $e^{4J} \ll \Phi_1^*/\Phi_2^*$, we expect a decrease of the critical concentration for small values of $1/\alpha$. On the other hand, if $e^{4J} \gg \Phi_1^*/\Phi_2^*$, the critical concentration increases upon a relative increase of the species 2 population.

\subsection{Bidisperse case for $\Phi \ll \Phi^*$}
If the total number density of monomers is sufficiently small, only monomers and dimers are present in the solution, and this corresponds to investigating the limit $z_i= \exp(\mu_i)\ll 1$. In this case, we can make a Taylor expansion of Eq. \ref{eq:sum1} and Eq. \ref{eq:sum2} in terms of fugacities $z_i$ up to the second order, i.e., considering only the contribution of free monomers and dimers. This gives
\begin{equation}
\label{eq:L1}
\Phi_{1}+\Phi_{2}\simeq z_1 \exp(a_{1})+z_2 \exp(a_{2})+ 4 \exp(b_{12})z_1z_2+2 \exp(b_{11})z_1^2+2 \exp(b_{22})z_2^2,
\end{equation}
in which the first two terms are the contribution of monomers, while the 3 other terms present the contribution of dimers of type 12, 11 and 22. Furthermore we have
\begin{equation}
\label{eq:L2}
\Phi_{1}-\Phi_{2}\simeq z_1 \exp(a_{1})-z_2 \exp(a_{2})+ \exp(b_{11})z_1^2- \exp(b_{22})z_2^2.
\end{equation}
Combining equations \ref{eq:L1} and \ref{eq:L2}, we find that $ \alpha \approx \alpha^f$.  The reason, of course, is that in the low concentration regime, the assemblies are mainly in the monomeric regime.

With these approximations, we can  solve for $z_1$ and $z_2$ and calculate the fraction of self-assemblies $f$ up to the lowest order in terms of total concentration as well as the ratio of the two components $\alpha=\Phi_1/\Phi_2$,
\begin{eqnarray}
\label{eq:fraction}
 f \simeq \frac{2\alpha \Phi}{(1+\alpha)^2}[ 2 \exp(b_{12}-a_{1}-a_{2})+\alpha \exp(b_{11}-2a_{1})+\alpha^{-1} \exp(b_{22}-2a_{2})].
\end{eqnarray}
Again, we find that the first dominant term is linear in the total concentration. Its slope, however, depends on the stoichiometric ratio of the two components as well as on the free-energy parameters involved in the system, where $f\rightarrow 0$ if the activation energies $a_i$ are large. Equation \ref{eq:fraction}  demonstrates that even at the level of dimer formation, the coupling between the two components cannot be ignored.  As before, we obtain similar results as those for the monodisperse case in the limits $\alpha\rightarrow \infty$ and $\alpha\rightarrow 0$, as one would expect. If $\alpha$ is large then $f$ should be proportional to $\Phi_1$ and if it is small then it should be proportional to $\Phi_2$.

Now that we understand the behavior of our system in the limits of low and high concentrations, in the next section we examine numerically the full concentration behavior of self-assembly in  the case of ``ferromagnetic'' or blocky copolymeric ordering, $J>0$.

\section{Fraction of self-assemblies and critical concentration}
\begin{figure}[h!]
\begin{center}
\includegraphics[scale=0.6]{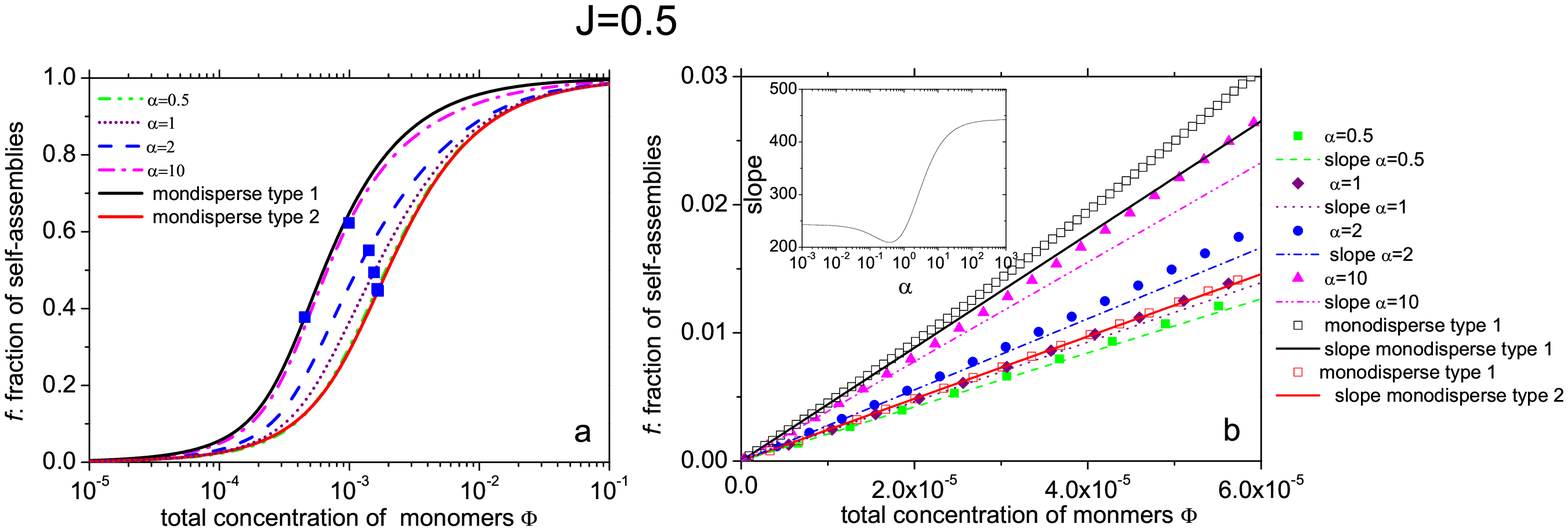}
\includegraphics[scale=0.58]{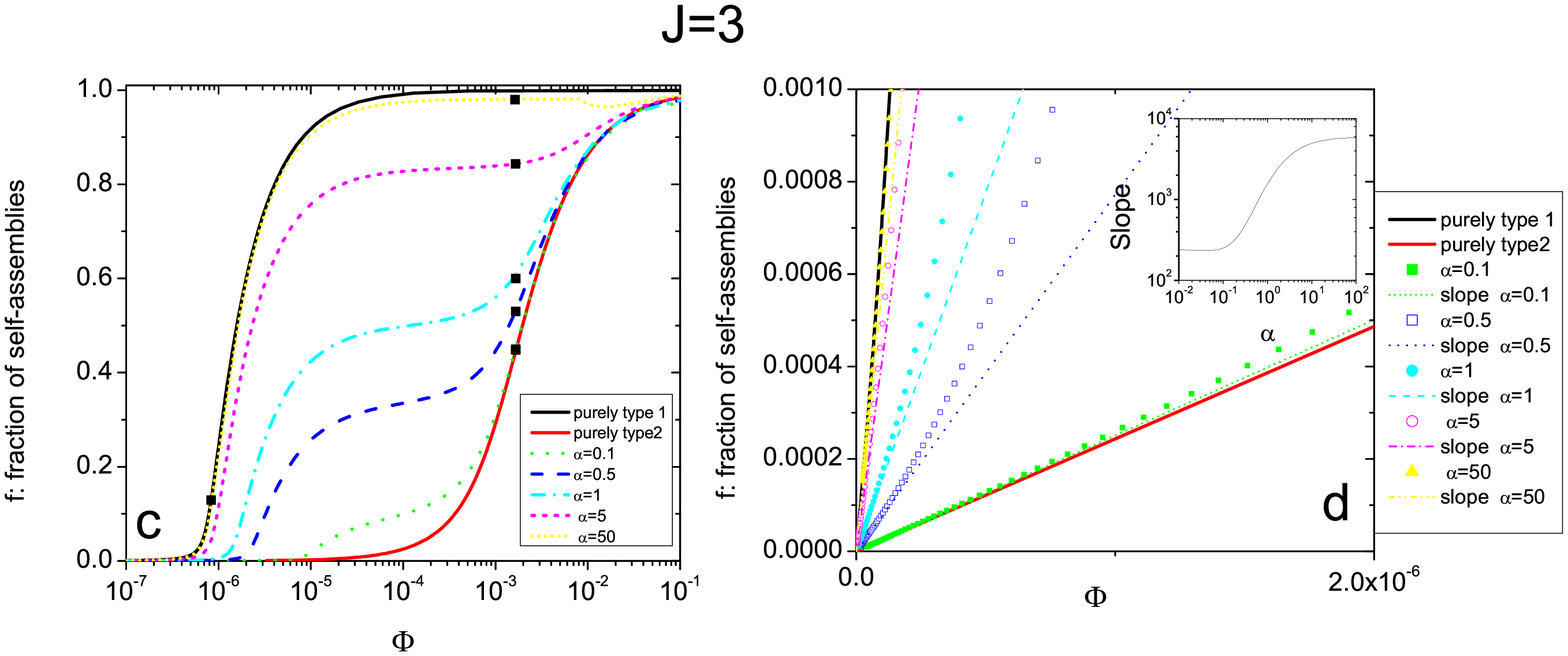}
\caption{The fraction of self-assemblies $f$ as a function of  the overall molar fraction of dissolved material $\Phi = \Phi_1+\Phi_2$ at different stoichiometric ratios $\alpha=\Phi_1/\Phi_2$ of the two components $1$ and $2$. Figures \ref{fig2}\emph{a} and \ref{fig2}\emph{b} show the results for weak bidispersity ($b_{11}=10$, $b_{12}=b_{22}=8$, $a_{1}=2.3$ and $a_{2}=1.6$). The squares on the curves in Fig. \ref{fig2}\emph{a} signify the fraction of  self-assemblies at the critical concentration, associated with each $\alpha$, as discussed in the main text. Figures \ref{fig2}\emph{c} and \ref{fig2}\emph{d} graphs show the case of a strong degree of bidispersity ($b_{11}=20$, $b_{12}=b_{22}=8$ , $a_{1}=6$ and $a_{2}=1.6$). The squares on each $\alpha$ curve in \ref{fig2}\emph{c} present  the fraction of  self-assemblies at their respective critical concentrations. The solid lines show the slopes for each $\alpha$ calculated based on the Eq. \ref{eq:fraction} and show a good agreement with numerical results at sufficiently low values of $\Phi$. In the insets, we depict the slope of $f$ at low concentrations as a function of $\alpha$ according to Eq. \ref{eq:fraction}.}

\label{fig2}
\end{center}
\end{figure}

We first focus on the dependence on the total concentration of the fraction of dissolved material present in self-assemblies, as we vary the ratio of the concentrations of the two components. In Fig. \ref{fig2}, we have plotted the fraction of self-assemblies $f$ as a function of the total molar fraction $\Phi$ for the particular case where we have assumed that $b_{12}=b_{22}$. Results for two cases are shown. In the first case, the  binding and activation free energies for the two species are slightly different (corresponding to weak bidispersity) and in the second the  binding and activation free energies for the two species are considerably different (corresponding to strong bidispersity). The chosen values of the activation and binding free energies in the first case correspond to oligopeptides consisting of 4 and 5 monomers and are sufficiently small to result in a relatively gradual self-assembly transition, while in the second case the free energy parameters of the first species chosen are typical values for an 11-mer \cite{Amalia-PNAS}.

First, we discuss the case of weak bidispersity depicted in figures \ref{fig2}a and  \ref{fig2}b. In this case, the fraction of self-assemblies increases gradually from zero to 1 for all ratios of the two components, $\alpha$, as one increases the total concentration of monomers. The bidisperse assembly curves are in between those corresponding to the two monodisperse cases. We find that for all the curves a crossover from an increasing slope for relatively low concentrations to a decreasing slope for very high concentrations can be observed, demonstrating a transition from minimal assembly to self-assembly dominated regime.  The activation free energies were chosen not to be very large, so we do not observe a sharp polymerization transition from 0 to a non-zero value.
In Fig.\ref{fig2}b, we have shown  the fraction of self-assemblies at low concentrations. To verify the validity of the low-concentration expansion, for each stoichiometry $\alpha$, we have also plotted the lines corresponding to  Eq. \ref{eq:fraction}. As can be seen, in this regime the fraction of assemblies $f$ grow linearly  as a function of concentration with the same slope predicted by  Eq. \ref{eq:fraction}. A careful look at this figure shows that the slope of $f$ versus $\alpha$ is non-monotonic. To highlight this, we have plotted the slope of $f$ versus $\alpha$ in the inset of Fig. \ref{fig2}b. The curves demonstrate the highly non-linear effects that are the result of mixing different kinds of assembler units.

Now, we turn to the case of strong bidispersity (large $J$) as depicted in Figs. \ref{fig2}c and \ref{fig2}d. Here, we also observe that the self-assembly curves for $f$ are between the curves of the pure species as we vary the ratio of the two components. However, the behavior of the mixed material as shown in the curves for the intermediate concentrations is dissimilar to that of the monodisperse solutions. We notice that the fraction of self-assemblies grows at low concentrations with a slope that depends on $\alpha$. Importantly, it grows linearly at very low $\Phi$ and in accord with our estimates for $\Phi \ll \Phi^*$ in Eq. \ref{eq:fraction}, as shown in  Fig. \ref{fig2}d.

However, after an initial relatively steep growth stage, an intermediate stage emerges for which the fraction of self-assemblies shows little variation with $\Phi$. Finally, for large enough concentrations the fraction of self-assemblies enters into a third regime of growth that follows the behavior predicted for high concentration region  $\Phi \gg \Phi^*$, i.e., deviating from  unity inversely proportional to $\Phi$.

As discussed in the previous section, in an assembly consisting of a single species of type $i$, the transition from mainly monomeric regime to an assembly-predominated regime can be demarcated by a critical concentration $\Phi_i^*\equiv \exp(-b_{ii}+a_i)$, which is equal to the maximum molar fraction of free monomers reached in the high-concentration regime. Indeed, for large enough activation free energies $a_i$, this corresponds to a sharp transition where the critical concentration identifies a  transition from no assembly to an  assembly-dominated state and $f(\Phi^*)\approx 0$. For small values of the activation free energy, where the transition is not sharp, the critical concentration signifies the crossover from the low concentration regime to the high concentration regime, and $f(\Phi^*)$ has a non-zero value close to 0.5 represented  by the squares in Fig. \ref{fig2}a, for the $f$ curves of pure assemblies of type $1$ and $2$. It would be  interesting to see if this critical concentration can also characterize the transition from minimal assembly to self-assembly dominated region in the mixture of two species and, if so, how it depends  on the stoichiometry $\alpha$.

In Fig. \ref{fig2}a and \ref{fig2}c, we have marked on each curve the points that correspond to the critical concentrations calculated according to Eqs. \ref{eq:C*}. We notice that in the weakly bidisperse curve, the value of $f$ at the critical concentration is around 0.5, therefore, one can think of $\Phi^*$ as demarcating the polymerization transition. However, in the strongly bidisperse case, the value of $f$ at $\Phi^*$ can have any value depending on $\alpha$. Particularly for large values of $\alpha$, we find $f$ to be close to 1. In these cases, the critical concentration seems to mark the onset of the transition from the plateau region in $f$ curves to the high-concentration regime, where $f \simeq 1- \Phi^*/ \Phi$.

\begin{figure}[h!]
\begin{center}
\includegraphics[scale=0.28]{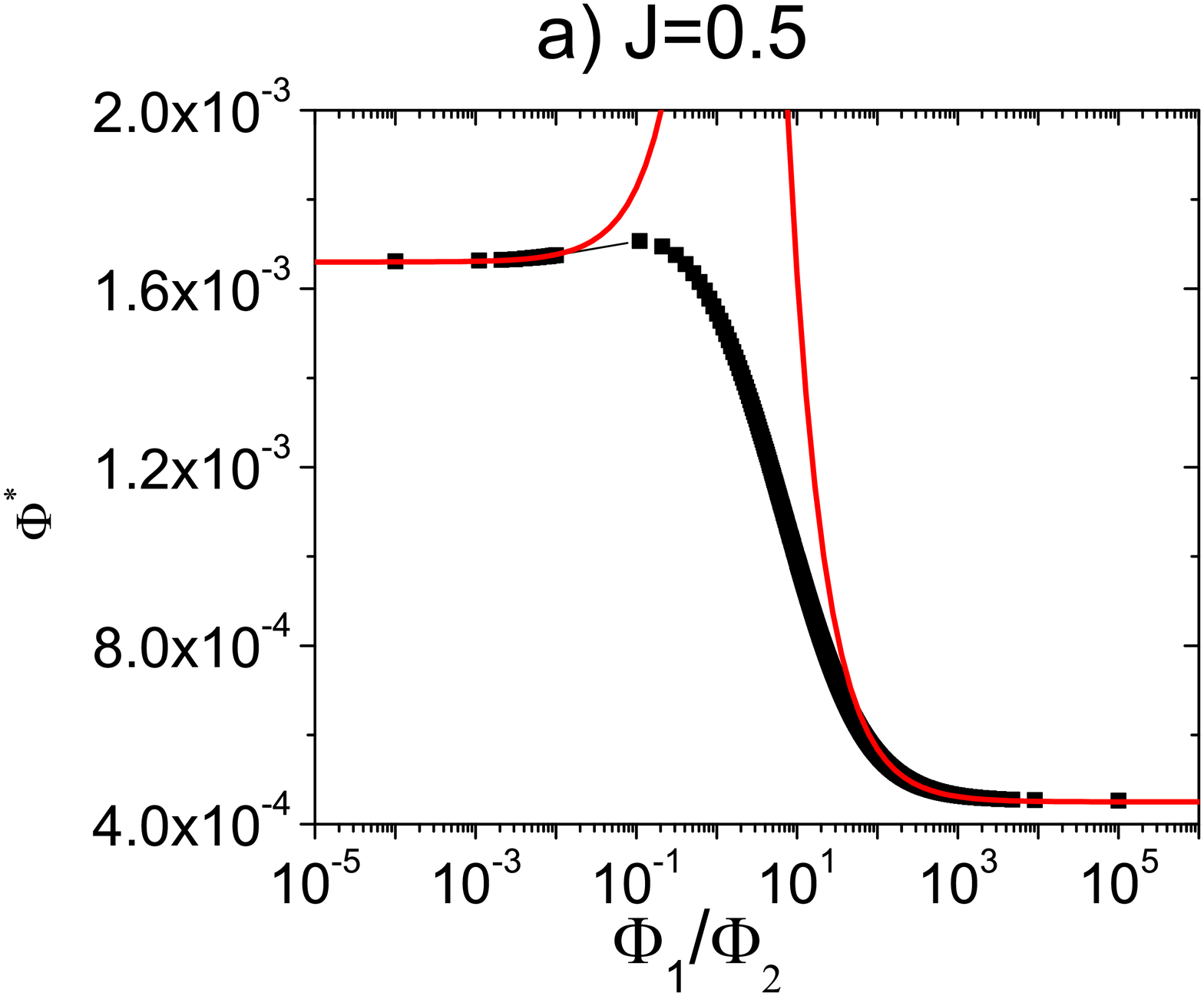}
\includegraphics[scale=0.27]{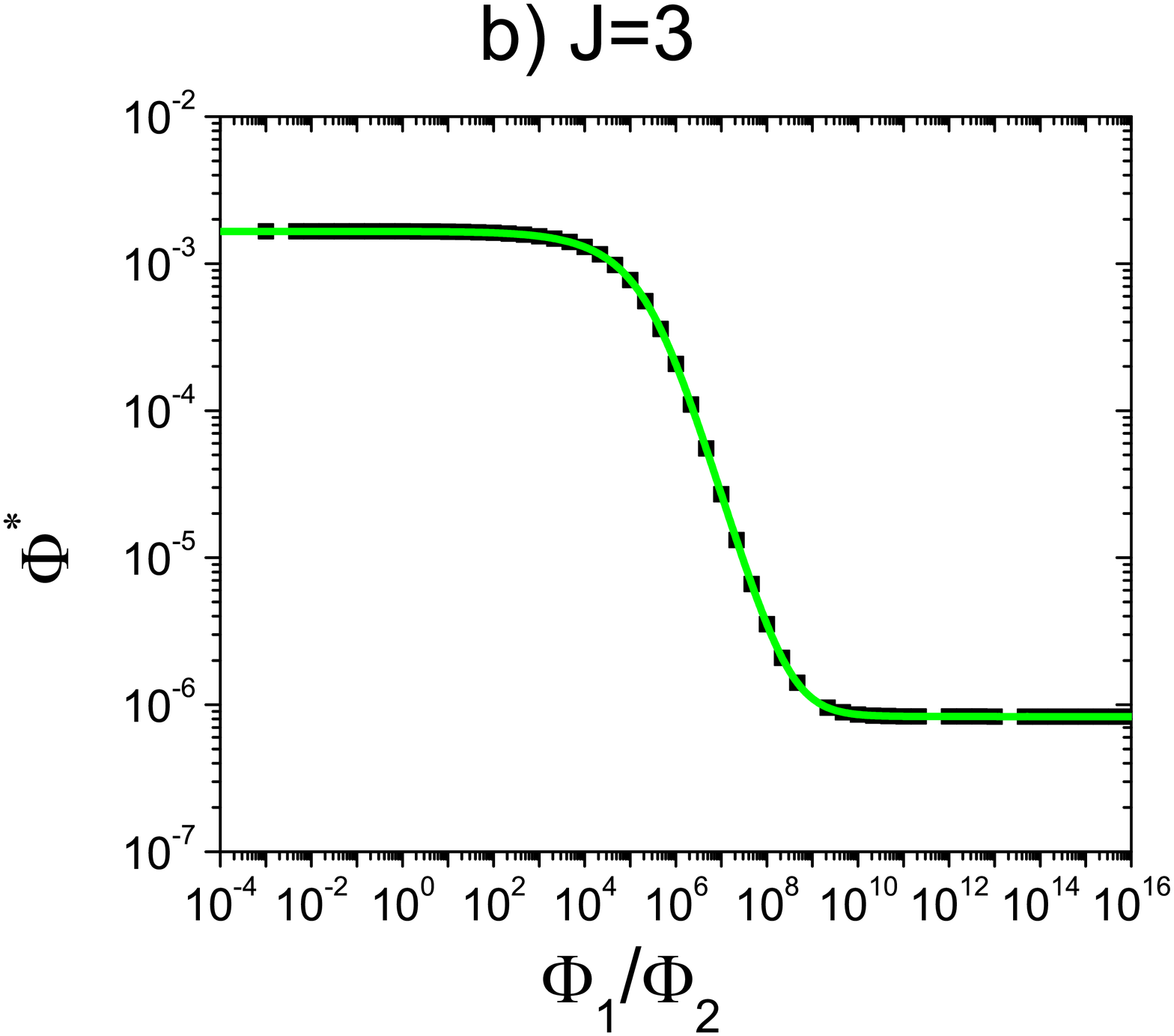}
\caption{  The critical molar fraction $\Phi^*$ as a function of ratio of the two components $\alpha=\Phi_1/\Phi_2$  plotted for a) weak bidispersity: $b_{11}=10$, $b_{12}=b_{22}=8$  , $a_{1}=2.3$ and $a_{2}=1.6$, corresponding to $J=0.5$  , $\Phi_1^*=4.528 \times 10^{-4}$ and  $\Phi_2^*=0.00166$ .  The lines show the approximate functions valid for very low and very high $\alpha$ values according to Eq. \ref{eq:C*1};  b) strong bidispersity: $b_{11}=20$, $b_{12}=b_{22}=8$ , $a_{1}=6$  and $a_{2}=1.6$  corresponding to a large value of $J=3$,  $\Phi_1^*=8.315 \times 10^{-7}$ and  $\Phi_2^*=0.00166$. The line shows the  analytical results obtained for the critical concentration in the large $J$ limit   based on  Eqs. \ref{eq:C*03}-\ref{eq:C*04}.  }
\label{fig3}
\end{center}
\end{figure}

Therefore, it would be interesting to obtain the full functional dependence of $\Phi^*$ on $\alpha$ and compare it to the results obtained in the limiting cases of $J \gg 1$, and  $ \alpha \ll e^{-4J}$ and $ \alpha \gg e^{4J}$ for any $J$. Here, we extract the full $\alpha$-dependence of $\Phi^*$ by solving the  Eqs. \ref{eq:C*} numerically for two sets of values of free energy parameters corresponding to small and large $J$ values as presented in Fig. \ref{fig3}a  (for $J=0.5$ $k_BT$) and  Fig. \ref{fig3}b (for $J=3$ $k_BT$). For the case of small $J$ in Fig. \ref{fig3}a, we have shown the results of our estimates of $\Phi^*$ for  $ \alpha \ll e^{-4J} $ and $ \alpha \gg e^{4J} $. We find  very good agreement with Eqs. \ref{eq:C*1}  for sufficiently small and large $\alpha$ values.

In Fig. \ref{fig3}b, on the other hand, we have depicted  the full functional dependence of the critical concentration on $\alpha$ obtained  in the limit of large $J$ based on  Eqs. \ref{eq:C*03}-\ref{eq:C*04}. Again we find very good agreement.
The general trend that we find is that the critical concentration value agrees with that of species 2 at small ratios  $\alpha=\Phi_1/\Phi_2$, and approaches the value of the critical concentration of species 1 for large enough $\alpha$ values, as it should be. For the small $J$ case, the crossover is not quite monotonic in $\alpha$, as is clear from Fig. \ref{fig3}a. The reason is that if $\alpha \ll 1$ and $e^{4J} > \Phi_2^*/\Phi_1^*$, as is the case here, we initially observe a slight increase of critical concentration  as discussed in the previous section.

The shown curves demonstrate that the  width of the region where the critical concentration deviates from that of either of the two species is a strong function of $J$, and the larger the asymmetry between the two species is, the wider is this region. More interestingly, for the case of strong asymmetry the critical value is identical to that of species 2 up to $\alpha \approx 10^2$. This probably means that  in this case the mixing of the two species in assemblies is not preferred in some intermediate regions, as we will discuss in a following section.

We can get an estimate of the width of  the critical concentration curve from the limiting formulas of Eq. \ref{eq:C*1} for the critical concentration. We can expect the deviation of the critical concentration from that of species 1 or 2, when the contributions of the terms proportional to $\Phi_1^*$ and $\Phi_2^*$ become equal. This gives us two onsets for low and high $\alpha$ values, $\alpha_L=\Phi_2^*/\Phi_1^* e^{-4J}$ and $\alpha_H=\Phi_2^*/\Phi_1^* e^{4J}$, leading to a width $2 \Phi_2^*/\Phi_1^* \sinh (4J)$. Inserting  the values of  the free energy parameters  for the curves shown in Fig. \ref{fig3}, we find  good agreement for the estimated width (results not shown).\\

In conclusion, both the fraction of assemblies and the critical concentration not only depend on the free energetic parameters and total concentration, but are also sensitive functions of the relative abundance of the two components. This implies that even mild contamination with a chemically distinct species potentially has a large effect on the degree of polymerization. In particular, having a critical concentration that depends on $\alpha$, the interesting  question that arises is how the fraction of the two components in the assemblies differs from $\alpha$. This is the subject of the following section.

\section{Distribution of monomers in  the assemblies}
Having a clear picture of the $\alpha$-dependence of the critical concentration,  $\Phi^*$, and the fraction of material in assemblies, $f$, our aim is to get a deeper insight into the composition of the self-assemblies that are formed. We would like to know how the two species are distributed in  the  assemblies, and whether the relative population of  the two species in each assembly is the same as that in bulk solution, i.e., $\alpha$. Of particular interest is whether or not each assembly is  formed of one type of species or if both species contribute to the formation of each assembly.

\begin{figure}[h!]
\begin{center}
\includegraphics[scale=0.28]{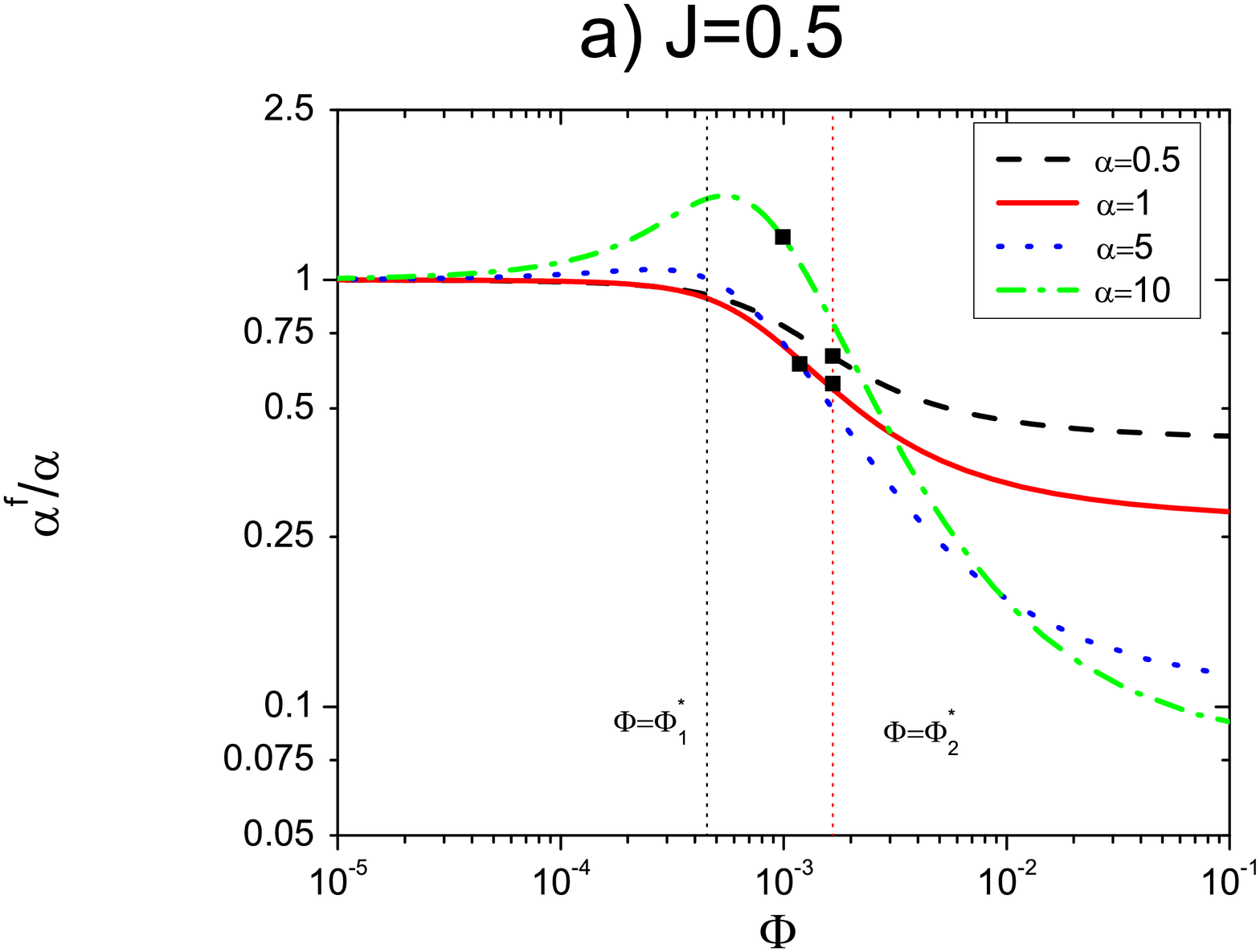}
\includegraphics[scale=0.25]{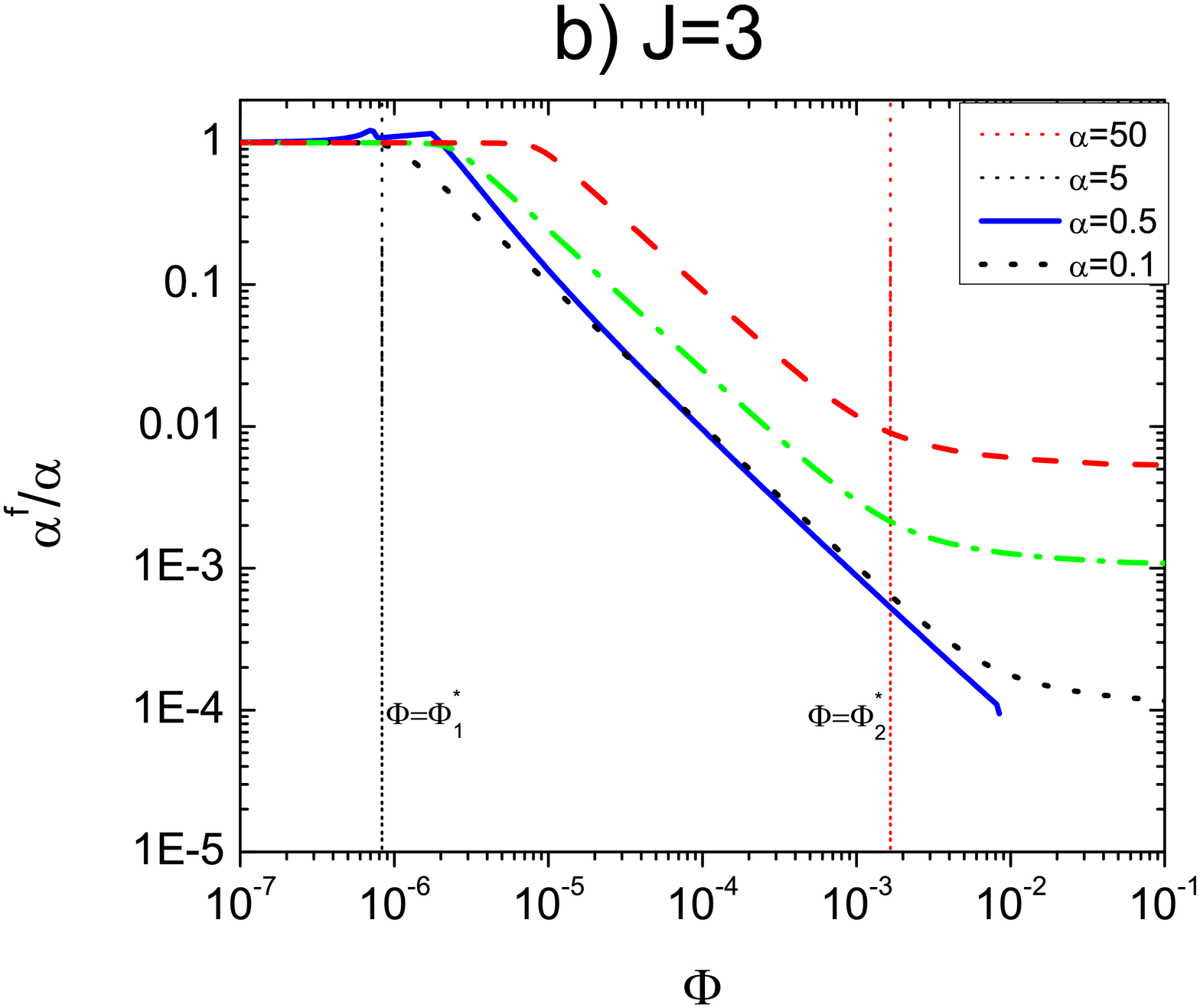}
\caption{ a) Ratio of the concentrations of the free monomer of species 1 and 2  $\alpha^f$ divided by the ratio of  total density of monomers present in the solution $\alpha$, as a function of  total concentration shown for different values of  $\alpha=\Phi_1/\Phi_2$ . The corresponding $\alpha$  values are depicted in the legends.  a)  Weak  bidispersity, with an equivalent  $J=0.5 $. b) strong bidispersity corresponding to  $J=3$ . The dotted lines correspond to concentrations $\Phi=\Phi_1^*$ and $\Phi=\Phi_2^*$, as indicated in the figure. The free energy parameters used here are the same as those of figures \ref{fig2} and \ref{fig3}.  }
\label{fig4}
\end{center}
\end{figure}

First, let us see if the ratio of the densities of the free monomers is conserved, as we increase  the total concentration of monomers and assemblies begin to form. In Fig. \ref{fig4} we have depicted the relative abundance of free monomers as a function of total concentration for different values of $\alpha$.  We find that for very low concentrations the density ratio of free monomers is the same as $\alpha$. However, for  larger $\Phi$ values the density ratio of free monomers drops and is considerably lower than $\alpha$. This implies that a larger fraction of  species of type 1  monomers (the species with greater tendency to self-assemble) contribute to the formation of self-assemblies.

The ratio of free and bound monomers keeps on decreasing with increasing concentration, until the total population of free monomers saturates, i.e., when $\Phi \gg \Phi^*$. Indeed, calculating the density of free monomers $\rho_i^f $ in the large $\alpha$ limit, we find that $\rho_{1}^f/\rho_{2}^f=\Phi_1^*/\Phi_2^* \exp(-4J)$. For large $\alpha$s, we observe an initial increase of the ratio of bound and free monomers in the vicinity of  $\Phi^{*}$, followed by a subsequent decrease, i.e., the dependence on $\alpha$ is non-monotonic. We note that $\alpha^f =( \Phi_1^f)/ (\Phi_2^f)$ will grow if $\Phi_2^f$ decreases. For $\alpha > 1$, especially if $\alpha$ is considerably larger than one,  there is a lack of  monomers of the second species in the system and if  $\Phi \approx \Phi_1^*$, the second  species also contributes to the assemblies made mainly of the first kind. This effect is  stronger if the difference in binding energies, and hence, $J$, is small.  Therefore,  in such a situation $(\Phi_2)^ f$ can become small and lead to  an increase of $\alpha^f$ around $\Phi \approx \Phi_1^*$.

We can also determine the average fraction of monomers of each species along assemblies of arbitrary length $N$. The fraction of monomers of species $j$ in a specific assembly of length $N$ is defined as  $\theta_{j}=(1/N) \sum_{i=1}^{N}n^{(j)}_i$. Note that $\theta_{j}$ in an assembly of length $N $  varies from one assembly to another according to $P_N(\theta_{j})$. However, we can calculate its average value $\langle \theta_{j} \rangle_N$ as a function of $N$ from the partition function $Z_N$, i.e., $\langle \theta_{j} \rangle_N=(1/N) \partial \ln Z_N/\partial \ln z_{j}$.
The constraint  $\sum_{i=1}^{N} (n^{(1)}_i+n^{(2)}_i)=N$ for each assembly of arbitrary length implies that $\langle \theta_{2} \rangle_N=1-\langle \theta_{1} \rangle_N$, therefore, from now on we focus on $\langle \theta_{1} \rangle_N$.

\begin{figure}[h!]
\begin{center}
\includegraphics[scale=0.8]{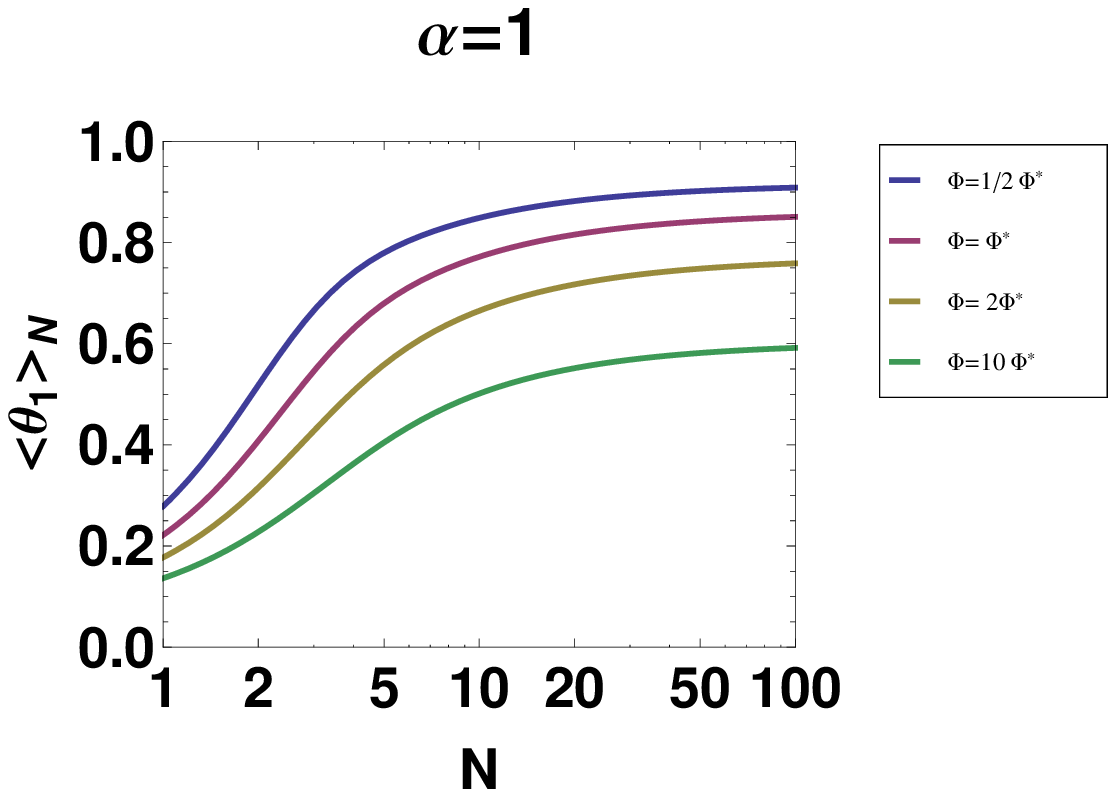}
\includegraphics[scale=0.65]{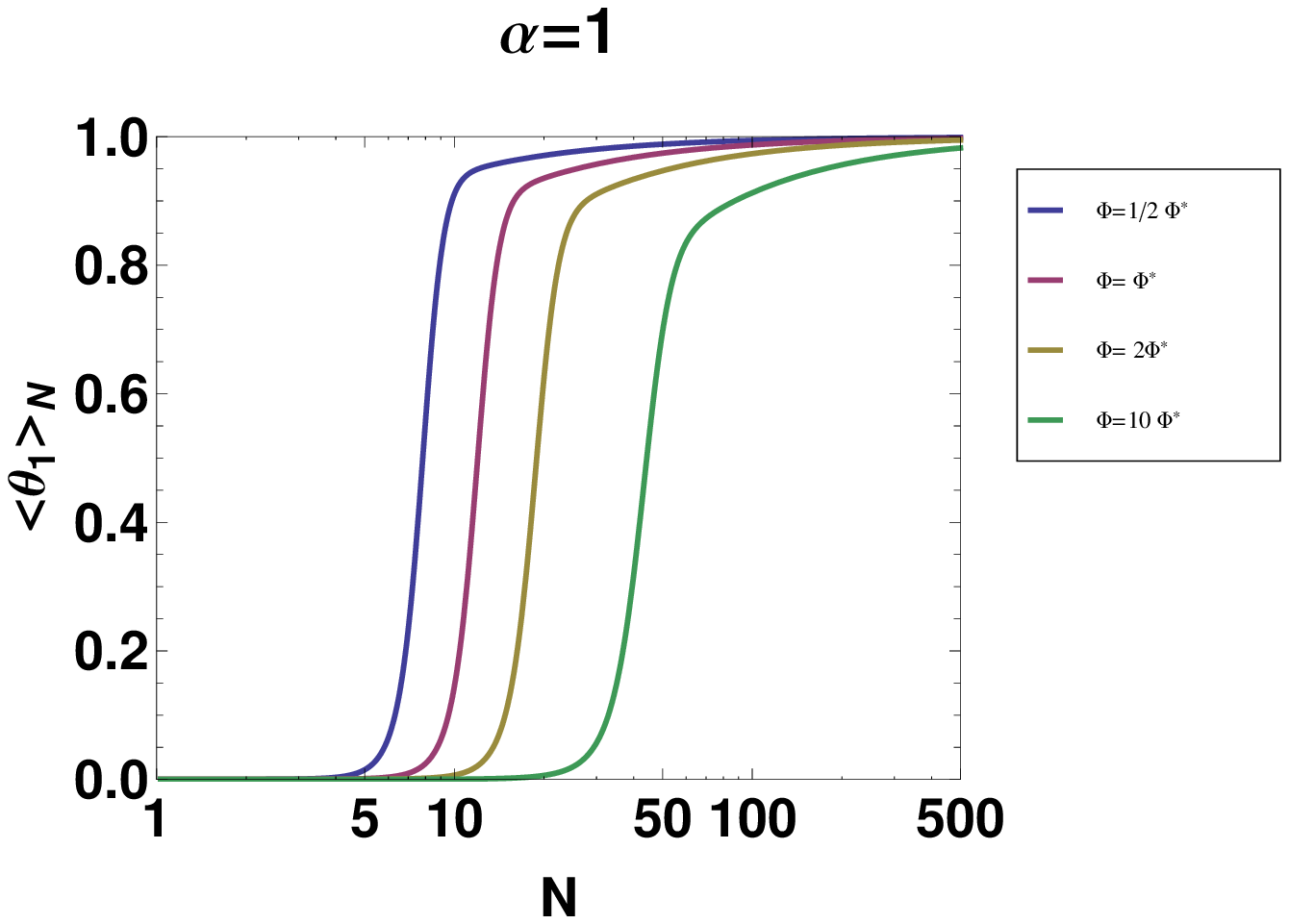}
\caption{The average fraction of monomers of type 1  for a composition of $\alpha=1$ as  a function of $N$ shown for different concentrations.  The upper graphs shows a weakly  bidisperse case, with an equivalent  $J=0.5 $ , while the lower graph shows a strongly bidisperse case, corresponding to  $J=3$ . The free energy parameters used here are the same as those of figures \ref{fig2} and \ref{fig3}. }
\label{fig5}
\end{center}
\end{figure}

In Fig. \ref{fig5}, we have plotted $\langle \theta_{1} \rangle_N$  for the special case of $\Phi_1=\Phi_2$, for both small and large $J$ values at several concentrations. This case is particularly illustrative, as deviations of  the fraction of the two species from 0.5 reflects the deviation of the distribution of the species from the original distribution, i.e., $\Phi_1=\Phi_2$. As Fig. \ref{fig5} indicates, for both small and large $J$ cases, the fraction of the species 1, and hence that of species 2, differs from one half. This figure clearly demonstrates that short assemblies are made mainly of species 2, i.e., the one with less tendency to self-assemble while the dominant population of long assemblies are species 1 (the species with a greater tendency with self-assembly).
The behavior  in the case of large $J$ is particularly interesting. We notice that at each concentration, assemblies shorter than  $ N_{c1}$ are made purely of type 2 species and  very long assemblies are made mostly of species type 1, and that the transition from compositionally pure assemblies of type 2 to compositionally pure assemblies of type 1 is sharp. This clearly demonstrates the ``demixing'' effects occurring due to the large asymmetry of the two species.

The behavior that we find is consistent with that obtained, in a more general sense, from the 1-D Ising model. This more general picture arising from the Ising model is that for very short chains, that is, compared to the correlation length $ \xi_0 \equiv \exp(2J)/2$, there is less combinatorial entropy available, simply because there is not enough room to move the domains about. As a result, states of either spin up or spin down (in our case assemblies merely consisting of type 1 or 2) are preferred.

In the other extreme of  chain lengths much longer than the correlation length $\xi_0$, the combinatorial factor can benefit from a large number of fragmentations, leading to many bound domains of moderate lengths of the order of the correlation length. For $J=0.5 k_BT$, $ \xi_0 \simeq 1.4$ while for $J=3 k_BT$, $ \xi_0 \simeq 202$. Therefore, for small $J$, almost for any assembly length  we are already in the regime that mixed assemblies are favored, while for large $J$, except for very long assemblies and sufficiently
high concentrations, we are in the region $N < \xi_0$, therefore, pure assemblies are preferred.\\

\begin{figure}[h!]
\begin{center}
\includegraphics[scale=0.27]{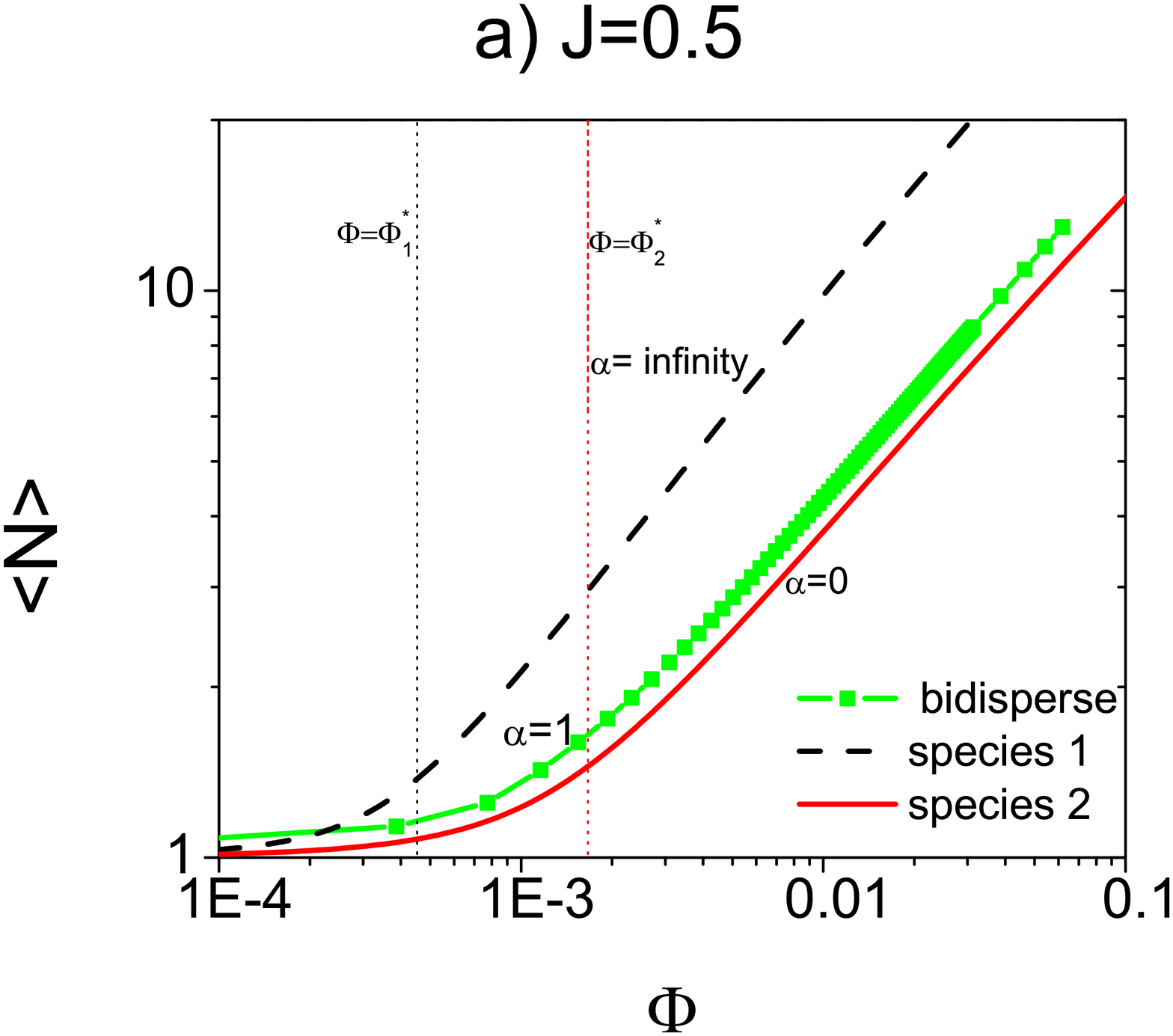}
\includegraphics[scale=0.255]{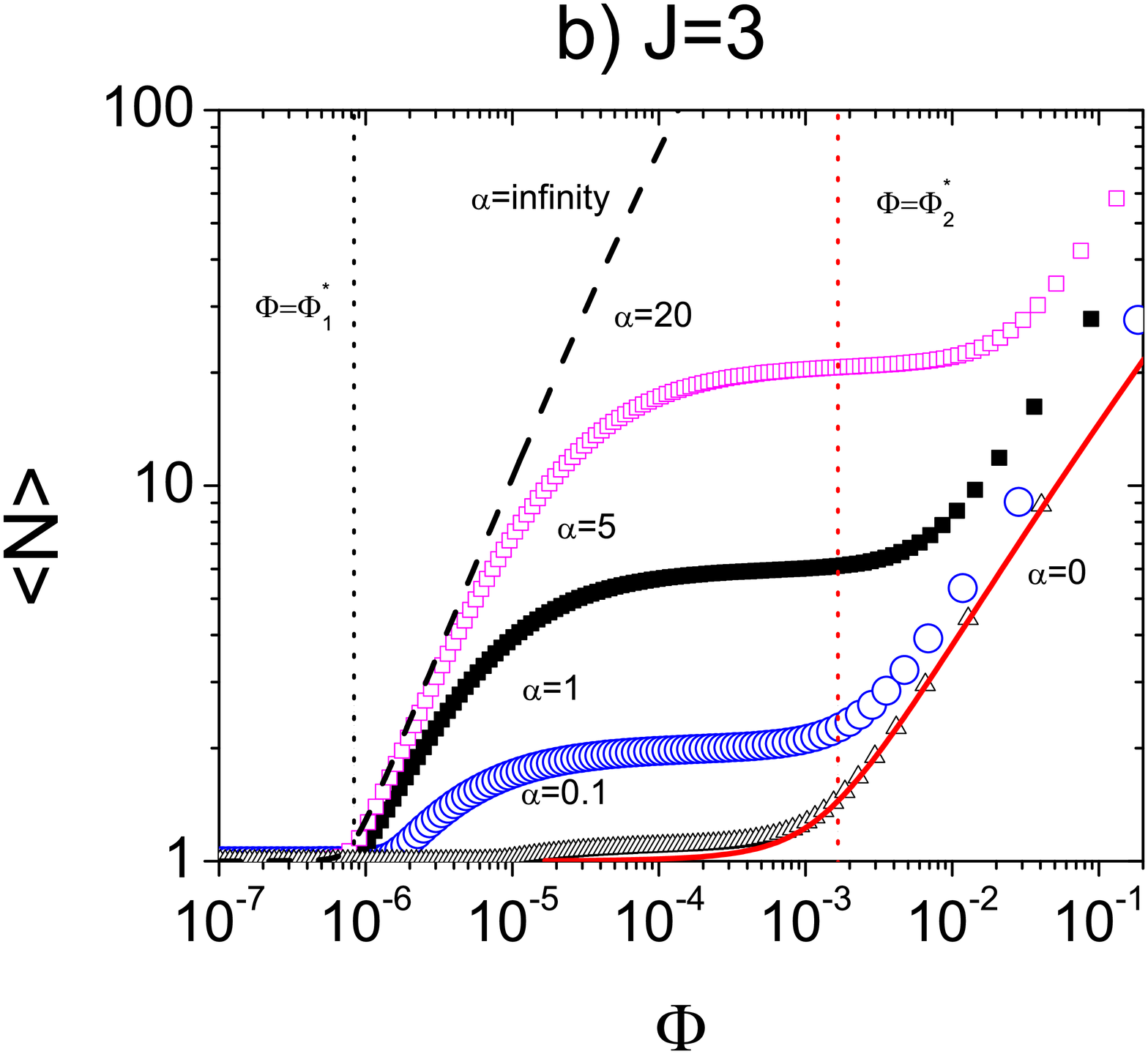}
\caption{ The number-averaged degree of polymerization of linear assemblies as a function of the overall concentration  for different stoichiometric ratios $\alpha$.  $\alpha=0$ and $\alpha=\infty$  correspond to monodisperse cases of type 1 and 2, respectively. a)  Weakly  bidisperse case, with an equivalent  $J=0.5 $ . b) Strongly bidisperse case corresponding to  $J=3$ . The free energy parameters used here are the same as those of Fig. \ref{fig2} and \ref{fig3}. The vertical dotted lines in each figure show the concentrations $\Phi=\Phi_1^*$ and $\Phi=\Phi_2^*$, respectively.}
\label{fig6}
\end{center}
\end{figure}

Having discussed  the average fraction  of the two species in  the assemblies $\langle \theta_{i} \rangle_N$, we next consider the number-averaged size of the assemblies, $\langle N\rangle$, as a function of  the total concentration of monomers. See Fig. \ref{fig6}. In this figure we also show the concentration dependence of $\langle N\rangle$  of  the monodisperse species for comparison. As expected, the average length of assemblies increases as we increase the total concentration of monomers. We find that the average length of the assemblies in the bidisperse case lies in between that of the pure species.
This implies that, e.g., for $\alpha=1$, the mean degree of polymerization obeys an apparent growth law that deviates from the usual square-root law over, say, a decade in concentration. While for small $J$, $\langle N\rangle$ grows monotonically with concentration, for the large-$J$ case $\langle N\rangle$ remains fairly constant at intermediate concentrations due to the demixing effect discussed earlier. This means that over a range of concentrations the mean aggregation number does not grow at all.

In this range of concentrations, the species 1 monomers have already formed relatively long assemblies, while the species 2  are still mainly in the form of free monomers. The average aggregation number,  $\langle N \rangle$,  reflects the average number of assemblies formed by \textit{both } species. Therefore, in the averaging, the growing number of long assemblies made of species 1 are taken with the growing number of very short assemblies of the species 2, and effectively $\langle N \rangle$ does not grow.  This behavior is seen even when the relative population of second species is small $\alpha=20$, this implies that impurities can strongly affect the growth of linear supramolecular assemblies. It shows again the important effect that impurities can have on observed quantities such as the average degree of polymerization.\\

\section{Concluding remarks  and outlook}
To summarize, we have set up a model that allows us to investigate theoretically the effects of molecular bidispersity in dilute solutions of quasi-linear, self-assembling objects. Mapping our model onto the 1-D Ising model, we find that different arrangements of the two species in the assemblies take place, depending on the relative values of the binding free energies involved in the binding. These morphologies correspond to ferromagnetic (``blocky''), anti-ferromagnetic (``alternating'') and paramagnetic-like (``random'') ordering of the two species in the polymeric assemblies. See Fig. 1.

Analyzing our model for the case of ferromagnetic-like (or blocky-type copolymeric) ordering, we find a range of interesting phenomena. The fraction of self-assemblies and the value of  the critical concentration, which quantifies the crossover to the high-concentration regime, not only depend on the free energy parameters of the two components in the system but also on the relative abundance of the two species. The degree of asymmetry of the two species, described by a  coupling constant $J$,  strongly influences the dependence of the  critical concentration on the density ratio of the two components.

The larger the value of the coupling parameter $J$, the wider the region over which the critical concentration is different from either that of the two species. Looking at the distribution of the monomers of each species, we find that a large asymmetry encoded by a large value of $J$ gives rise to a larger asymmetry in the composition of the assemblies formed. For  sufficiently large values of $J$ (signifying a large degree of molecular asymmetry) this leads to the emergence of a demixing region, where pure assemblies, made up almost entirely of  either of the two species, coexist.

It is of interest, we reckon, to discuss the analogy of the effective coupling constant $J$ in our bidisperse model, and the Flory-Huggins binary interaction parameter $\chi$, appearing in polymer solutions and mixtures \cite{Flory}. In a polymer solution, the interaction parameter $\chi$ determines the miscibility of the solvent and the solute. Apparently, the parameters $J$ and $\chi$ govern ordering processes in mixtures, in the former that in supramolecular polymers and in the latter polymer solutions or blends. In our model, increasing $J$ leads to the segregation of species into compositionally pure assemblies, which is analogous to having a large $\chi$ between binary polymer mixtures leading to macroscopic phase separation \cite{Rubinstein}.

It is sensible comparing the results of our study with that of the existing phenomenology of  polydispersity effects on the phase behavior of thermodynamic systems. Commonly, the introduction of polydispersity causes a range of new features in the phase behavior. Two important observed effects of polydispersity are: i) A strong widening of the coexistence region, i.e., of the density range within which two or more phases coexist; and ii) fractionation, meaning that the coexisting phases have different concentrations to the different particle species present \cite{Sollich}.

Here, for the crossover from a mainly monomeric regime to a self-assembly dominated regime, characterized by a critical concentration,  we find a strong dependence of the critical concentration on the ratio of the two concentrations. Furthermore, we  observe an analogous effect to fractionation. The relative composition of the two species in the assemblies and hence also that of the free, unbound monomers in solution is different from the original parent ratio of monomers put into the system, especially for concentrations larger than the critical concentration.

Finally, although we have investigated the effect of polydispersity for the simplified case of  bidisperse solutions, we believe that the insights obtained provide  a general insight into the  general phenomenology and the qualitative features of more general assembly behavior in self-assembling systems. In the general polydisperse case, one would expect  the value of the critical concentration, which demarcates the transition from the monomeric regime to that where self-assemblies predominate, to depend on the exact form  of the distribution of the polydispersity attribute, and to be strongly affected by the width of the distribution.

One may deduce that for a weakly polydisperse system, i.e., one with a narrow distribution of the pertinent polydisperse attribute, a monotonic crossover from the monomeric regime to the self-assembly regime can be expected. For a strongly polydisperse system, characterized by a wide distribution function, the appearance of multiple regions where pure assemblies of single species coexist should be expected. Moreover, from our calculations we conclude that independent of the degree of polydispersity,  monomers with lower tendency to self-assemble are found more abundantly in shorter assemblies, while those with a larger affinity to self-assembly have a greater contribution to longer assemblies.

A model to investigate the generalized case of a continuous distribution of polydisperse attributes, similar to  that of the polydisperse  lattice gas model for the liquid-vapor phase equilibria, \cite{Sollich1},  and its consequences on self-assembly is,  currently under development \cite{Sollich2}.

\section*{Acknowledgment}
This work was part of the Research Programme of the Dutch Polymer Institute (DPI), Eindhoven, The Netherlands, as Project number 610. S. J-F would like to thank the foundation of ``Triangle de la Physique'' for further support with this project. We would like to thank Amalia Aggeli, Sarah Harris, Cor Koning, Bin Bin Liu, Tom McLeish  and Peter Sollich for stimulating discussions. \\

\newpage

\textbf{\large{List of symbols}}\\
\begin{itemize}
\item $b_{ij} $: free energy   of the bonded interaction  between two monomers of type $i=1,2$ and $j=1,2$, in units of the thermal energy $k_B T$;
\item $a_i $: activation free energy  of species $i=1,2$ in units of $k_B T$, with $K_{a i}$ defining the activation constant;
\item $n^{(j)}_i$ the occupation number of species of type $j=1,2$ at site $i$ of a specific assembly;
\item $\Omega$: grand potential energy in units of $k_B T$;
 \item $Z_{N}$: the partition function of an assembly of length $N$;
\item $\rho(N)$: the number density of assemblies of degree of polymerization $N$;
\item $\nu$: interaction volume;
\item $\langle N \rangle $: the number-averaged degree  of polymerization $N$;
\item $J\equiv\frac{1}{4}(b_{11}+b_{22}-2b_{12})$: The effective coupling constant in the Ising mapping;
 \item  $H\equiv \frac{1}{2}[(b_{11}-b_{22})+(\mu_1-\mu_2)]$: the magnetic field in the Ising model;
 \item $\bar{b}\equiv\frac{1}{4}(b_{11}+b_{22}+2b_{12})$: average binding free energy;
 \item $\mu_i$: chemical potential of species $i=1,2$ ;
 \item $z_i \equiv \exp(\mu_i)$: The corresponding fugacities of species $i$;
 \item $\lambda_\pm$: the eigenvalues of  the transfer matrix of Ising model;
\item $\Lambda_\pm \equiv \lambda_\pm \exp(\bar{b})\sqrt{z_1 z_2}$: The effective fugacities of the bidisperse system;
\item $\Phi_i$: molar fraction of  molecules of species $i=1,2$;
\item $\Phi \equiv \Phi_1+\Phi_2$: the overall molar fraction;
\item $\alpha \equiv \Phi_1/ \Phi_2$: the ratio of molar fraction of the two species ;
\item $f$: mean fraction of self-assemblies;
\item $\Phi_i^*= \exp(-b_i+a_i)$: critical molar fraction associated with species $i$, demarcating the transition from minimal assembly to assembly-predominated regime;
\item $\Phi^*(\alpha)$: critical molar fraction of the bidisperse system, whose value depends on the ratio of molar fraction of the two components;
        \item $\Phi^f_i$: the molar fraction of free monomers of species $i$ ;
        \item $\rho^f \equiv \Phi^f_1/ \Phi^f_2$: the  ratio of the molar fraction of free monomers;
\item$\theta_{j}=1/N \sum_{i=1}^{N}n^{(j)}_i$: the fraction of species of type $j=1,2$ in a specific  assembly of length $N$ ;
     \item $\langle \theta_{j} \rangle_N$: the average fraction of species of type $j=1,2$ in the collection of assemblies of length $N$;
\end{itemize}

\bibliography{bidisperse1}

\begin{thebibliography}{42}
\expandafter\ifx\csname natexlab\endcsname\relax\def\natexlab#1{#1}\fi
\expandafter\ifx\csname bibnamefont\endcsname\relax
  \def\bibnamefont#1{#1}\fi
\expandafter\ifx\csname bibfnamefont\endcsname\relax
  \def\bibfnamefont#1{#1}\fi
\expandafter\ifx\csname citenamefont\endcsname\relax
  \def\citenamefont#1{#1}\fi
\expandafter\ifx\csname url\endcsname\relax
  \def\url#1{\texttt{#1}}\fi
\expandafter\ifx\csname urlprefix\endcsname\relax\def\urlprefix{URL }\fi
\providecommand{\bibinfo}[2]{#2}
\providecommand{\eprint}[2][]{\url{#2}}

\bibitem[{\citenamefont{Holder and Sommerdijk}(2011)}]{Nico1}
\bibinfo{author}{\bibfnamefont{S.~J.} \bibnamefont{Holder}} \bibnamefont{and}
  \bibinfo{author}{\bibfnamefont{N.~A. J.~M.} \bibnamefont{Sommerdijk}},
  \bibinfo{journal}{Polym. Chem.} \textbf{\bibinfo{volume}{2}},
  \bibinfo{pages}{1018} (\bibinfo{year}{2011}).

\bibitem[{\citenamefont{Sommerdijk and de~With}(2008)}]{Nico2}
\bibinfo{author}{\bibfnamefont{N.~A. J.~M.} \bibnamefont{Sommerdijk}}
  \bibnamefont{and} \bibinfo{author}{\bibfnamefont{G.}~\bibnamefont{de~With}},
  \bibinfo{journal}{Chem. Rev.} \textbf{\bibinfo{volume}{108}},
  \bibinfo{pages}{4499–4550} (\bibinfo{year}{2008}).

\bibitem[{\citenamefont{McPherson}(2005)}]{McPherson}
\bibinfo{author}{\bibfnamefont{A.}~\bibnamefont{McPherson}},
  \bibinfo{journal}{BioEssays} \textbf{\bibinfo{volume}{27}}
  (\bibinfo{year}{2005}).

\bibitem[{\citenamefont{Lehn}(2002)}]{Lehn}
\bibinfo{author}{\bibfnamefont{J.-M.} \bibnamefont{Lehn}},
  \bibinfo{journal}{Polym. Int.} \textbf{\bibinfo{volume}{51}},
  \bibinfo{pages}{825} (\bibinfo{year}{2002}).

\bibitem[{\citenamefont{de~Greef et~al.}(2009)\citenamefont{de~Greef, Smulders,
  Wolffs, Schenning, Sijbesma, and Meijer}}]{de-Greef}
\bibinfo{author}{\bibfnamefont{T.~F.~A.} \bibnamefont{de~Greef}},
  \bibinfo{author}{\bibfnamefont{M.~M.~J.} \bibnamefont{Smulders}},
  \bibinfo{author}{\bibfnamefont{M.}~\bibnamefont{Wolffs}},
  \bibinfo{author}{\bibfnamefont{A.}~\bibnamefont{Schenning}},
  \bibinfo{author}{\bibfnamefont{R.~P.} \bibnamefont{Sijbesma}},
  \bibnamefont{and} \bibinfo{author}{\bibfnamefont{E.~W.}
  \bibnamefont{Meijer}}, \bibinfo{journal}{Chem. Rev.}
  \textbf{\bibinfo{volume}{109}}, \bibinfo{pages}{5687–5754}
  (\bibinfo{year}{2009}).

\bibitem[{\citenamefont{Koopmans and Aggeli}(2010)}]{Koopmans}
\bibinfo{author}{\bibfnamefont{R.}~\bibnamefont{Koopmans}} \bibnamefont{and}
  \bibinfo{author}{\bibfnamefont{A.}~\bibnamefont{Aggeli}},
  \bibinfo{journal}{Current Opinion in Microbiology}
  \textbf{\bibinfo{volume}{13}}, \bibinfo{pages}{327} (\bibinfo{year}{2010}).

\bibitem[{\citenamefont{Ciferri}(2005)}]{Ciferri}
\bibinfo{editor}{\bibfnamefont{A.}~\bibnamefont{Ciferri}}, ed.,
  \emph{\bibinfo{title}{Supramolecular Polymers}} (\bibinfo{publisher}{CRC
  Press}, \bibinfo{year}{2005}), \bibinfo{edition}{2nd} ed.

\bibitem[{\citenamefont{Weiss and Terech}(2006)}]{Weiss}
\bibinfo{author}{\bibfnamefont{R.~G.} \bibnamefont{Weiss}} \bibnamefont{and}
  \bibinfo{author}{\bibfnamefont{P.}~\bibnamefont{Terech}},
  \emph{\bibinfo{title}{Molecular gels: materials with self-assembled fibrillar
  networks}} (\bibinfo{publisher}{Springer}, \bibinfo{year}{2006}).

\bibitem[{\citenamefont{Koopmans}(2009)}]{Koop}
\bibinfo{editor}{\bibfnamefont{R.}~\bibnamefont{Koopmans}}, ed.,
  \emph{\bibinfo{title}{Advances in Chemical Engineering: Engineering Aspects
  of Self-Organising Materials}} (\bibinfo{publisher}{Academic},
  \bibinfo{year}{2009}).

\bibitem[{\citenamefont{Ariga et~al.}({2008})\citenamefont{Ariga, Hill, Lee,
  Vinu, Charvet, and Acharya}}]{SupraSA1}
\bibinfo{author}{\bibfnamefont{K.}~\bibnamefont{Ariga}},
  \bibinfo{author}{\bibfnamefont{J.~P.} \bibnamefont{Hill}},
  \bibinfo{author}{\bibfnamefont{M.~V.} \bibnamefont{Lee}},
  \bibinfo{author}{\bibfnamefont{A.}~\bibnamefont{Vinu}},
  \bibinfo{author}{\bibfnamefont{A.}~\bibnamefont{Charvet}}, \bibnamefont{and}
  \bibinfo{author}{\bibfnamefont{S.}~\bibnamefont{Acharya}},
  \bibinfo{journal}{Sci. Technol. Adv. Mater.} \textbf{\bibinfo{volume}{{9}}},
  \bibinfo{pages}{014109} (\bibinfo{year}{{2008}}).

\bibitem[{\citenamefont{Rothemund}({2006})}]{DNA-origami}
\bibinfo{author}{\bibfnamefont{P.~W.~K.} \bibnamefont{Rothemund}},
  \bibinfo{journal}{Nature} \textbf{\bibinfo{volume}{{440}}},
  \bibinfo{pages}{297} (\bibinfo{year}{{2006}}).

\bibitem[{\citenamefont{Aggeli et~al.}(2001)\citenamefont{Aggeli, Nyrkova,
  Bell, Harding, Carrick, McLeish, Semenov, and Boden}}]{Amalia-PNAS}
\bibinfo{author}{\bibfnamefont{A.}~\bibnamefont{Aggeli}},
  \bibinfo{author}{\bibfnamefont{I.~A.} \bibnamefont{Nyrkova}},
  \bibinfo{author}{\bibfnamefont{M.}~\bibnamefont{Bell}},
  \bibinfo{author}{\bibfnamefont{R.}~\bibnamefont{Harding}},
  \bibinfo{author}{\bibfnamefont{L.}~\bibnamefont{Carrick}},
  \bibinfo{author}{\bibfnamefont{T.~C.~B.} \bibnamefont{McLeish}},
  \bibinfo{author}{\bibfnamefont{A.~N.} \bibnamefont{Semenov}},
  \bibnamefont{and} \bibinfo{author}{\bibfnamefont{N.}~\bibnamefont{Boden}},
  \bibinfo{journal}{PNAS} \textbf{\bibinfo{volume}{98}}, \bibinfo{pages}{11857}
  (\bibinfo{year}{2001}).

\bibitem[{\citenamefont{Martin}(1996)}]{EP1}
\bibinfo{author}{\bibfnamefont{R.}~\bibnamefont{Martin}},
  \bibinfo{journal}{Chem Rev} \textbf{\bibinfo{volume}{96}},
  \bibinfo{pages}{3043–3064} (\bibinfo{year}{1996}).

\bibitem[{\citenamefont{Brunsveld et~al.}(2001)\citenamefont{Brunsveld, Folmer,
  Meijer, and Sijbesma}}]{EP2}
\bibinfo{author}{\bibfnamefont{L.}~\bibnamefont{Brunsveld}},
  \bibinfo{author}{\bibfnamefont{B.~J.~B.} \bibnamefont{Folmer}},
  \bibinfo{author}{\bibfnamefont{E.}~\bibnamefont{Meijer}}, \bibnamefont{and}
  \bibinfo{author}{\bibfnamefont{R.~P.} \bibnamefont{Sijbesma}},
  \bibinfo{journal}{Chem Rev} \textbf{\bibinfo{volume}{101}},
  \bibinfo{pages}{4071} (\bibinfo{year}{2001}).

\bibitem[{\citenamefont{Zhao and Moore}(2003)}]{EP3}
\bibinfo{author}{\bibfnamefont{D.}~\bibnamefont{Zhao}} \bibnamefont{and}
  \bibinfo{author}{\bibfnamefont{J.~S.} \bibnamefont{Moore}},
  \bibinfo{journal}{Org Biomol Chem} \textbf{\bibinfo{volume}{1}},
  \bibinfo{pages}{3471} (\bibinfo{year}{2003}).

\bibitem[{\citenamefont{Bouteiller}(2007)}]{EP4}
\bibinfo{author}{\bibfnamefont{L.}~\bibnamefont{Bouteiller}},
  \bibinfo{journal}{Adv Polym Sci} \textbf{\bibinfo{volume}{207}},
  \bibinfo{pages}{79} (\bibinfo{year}{2007}).

\bibitem[{\citenamefont{Paramonov et~al.}(2005)\citenamefont{Paramonov, Gauba,
  and Hartgerink}}]{polydisperse}
\bibinfo{author}{\bibfnamefont{S.~E.} \bibnamefont{Paramonov}},
  \bibinfo{author}{\bibfnamefont{V.}~\bibnamefont{Gauba}}, \bibnamefont{and}
  \bibinfo{author}{\bibfnamefont{J.~D.} \bibnamefont{Hartgerink}},
  \bibinfo{journal}{Macromolecules} \textbf{\bibinfo{volume}{38}},
  \bibinfo{pages}{7555–7561} (\bibinfo{year}{2005}).

\bibitem[{\citenamefont{Ryu et~al.}(2004)\citenamefont{Ryu, Oh, Zin, and
  Lee}}]{polydisperse1}
\bibinfo{author}{\bibfnamefont{J.-H.} \bibnamefont{Ryu}},
  \bibinfo{author}{\bibfnamefont{N.-K.} \bibnamefont{Oh}},
  \bibinfo{author}{\bibfnamefont{W.-C.} \bibnamefont{Zin}}, \bibnamefont{and}
  \bibinfo{author}{\bibfnamefont{M.}~\bibnamefont{Lee}}, \bibinfo{journal}{J.
  AM. CHEM. SOC.} \textbf{\bibinfo{volume}{126}}, \bibinfo{pages}{3551}
  (\bibinfo{year}{2004}).

\bibitem[{\citenamefont{Leibler and Benoit}(1981)}]{diblock1}
\bibinfo{author}{\bibfnamefont{L.}~\bibnamefont{Leibler}} \bibnamefont{and}
  \bibinfo{author}{\bibfnamefont{H.}~\bibnamefont{Benoit}},
  \bibinfo{journal}{Polymer} \textbf{\bibinfo{volume}{22}},
  \bibinfo{pages}{195} (\bibinfo{year}{1981}).

\bibitem[{\citenamefont{Hong and Noolandi}(1984)}]{diblock2}
\bibinfo{author}{\bibfnamefont{K.~M.} \bibnamefont{Hong}} \bibnamefont{and}
  \bibinfo{author}{\bibfnamefont{J.}~\bibnamefont{Noolandi}},
  \bibinfo{journal}{Polym. Commun.} \textbf{\bibinfo{volume}{25}},
  \bibinfo{pages}{265} (\bibinfo{year}{1984}).

\bibitem[{\citenamefont{Milner et~al.}(1989)\citenamefont{Milner, Witten, and
  Cates}}]{diblock3}
\bibinfo{author}{\bibfnamefont{S.~T.} \bibnamefont{Milner}},
  \bibinfo{author}{\bibfnamefont{T.~A.} \bibnamefont{Witten}},
  \bibnamefont{and} \bibinfo{author}{\bibfnamefont{M.~E.} \bibnamefont{Cates}},
  \bibinfo{journal}{Macromolecules} \textbf{\bibinfo{volume}{22}},
  \bibinfo{pages}{853} (\bibinfo{year}{1989}).

\bibitem[{\citenamefont{Burger et~al.}(1990)\citenamefont{Burger, Ruland, and
  Semenov}}]{diblock4}
\bibinfo{author}{\bibfnamefont{C.}~\bibnamefont{Burger}},
  \bibinfo{author}{\bibfnamefont{W.}~\bibnamefont{Ruland}}, \bibnamefont{and}
  \bibinfo{author}{\bibfnamefont{A.~N.} \bibnamefont{Semenov}},
  \bibinfo{journal}{Macromolecules} \textbf{\bibinfo{volume}{23}},
  \bibinfo{pages}{3339} (\bibinfo{year}{1990}).

\bibitem[{\citenamefont{van Gestel et~al.}(2004)\citenamefont{van Gestel,
  van~der Schoot, and Michels}}]{Gestel}
\bibinfo{author}{\bibfnamefont{J.}~\bibnamefont{van Gestel}},
  \bibinfo{author}{\bibfnamefont{P.}~\bibnamefont{van~der Schoot}},
  \bibnamefont{and} \bibinfo{author}{\bibfnamefont{M.}~\bibnamefont{Michels}},
  \bibinfo{journal}{J. Chem. Phys.} \textbf{\bibinfo{volume}{120}},
  \bibinfo{pages}{8253} (\bibinfo{year}{2004}).

\bibitem[{\citenamefont{Markvoort et~al.}(2011)\citenamefont{Markvoort, ten
  Eikelder, Hilbers, de~Greef, and Meijer}}]{Markvoort}
\bibinfo{author}{\bibfnamefont{H.~A.~J.} \bibnamefont{Markvoort}},
  \bibinfo{author}{\bibfnamefont{M.}~\bibnamefont{ten Eikelder}},
  \bibinfo{author}{\bibfnamefont{P.~A.} \bibnamefont{Hilbers}},
  \bibinfo{author}{\bibfnamefont{T.~F.} \bibnamefont{de~Greef}},
  \bibnamefont{and} \bibinfo{author}{\bibfnamefont{E.}~\bibnamefont{Meijer}},
  \bibinfo{journal}{Nature Communications} \textbf{\bibinfo{volume}{2}},
  \bibinfo{pages}{509} (\bibinfo{year}{2011}).

\bibitem[{\citenamefont{Smulders et~al.}(2011)\citenamefont{Smulders,
  Nieuwenhuizen, Grossman, Filot, Lee, de~Greef, Schenning, Palmans, and
  Meijer}}]{Smulders2}
\bibinfo{author}{\bibfnamefont{M.~M.~J.} \bibnamefont{Smulders}},
  \bibinfo{author}{\bibfnamefont{M.~M.~L.} \bibnamefont{Nieuwenhuizen}},
  \bibinfo{author}{\bibfnamefont{M.}~\bibnamefont{Grossman}},
  \bibinfo{author}{\bibfnamefont{I.~A.~W.} \bibnamefont{Filot}},
  \bibinfo{author}{\bibfnamefont{C.~C.} \bibnamefont{Lee}},
  \bibinfo{author}{\bibfnamefont{T.~F.~A.} \bibnamefont{de~Greef}},
  \bibinfo{author}{\bibfnamefont{A.~P. H.~J.} \bibnamefont{Schenning}},
  \bibinfo{author}{\bibfnamefont{A.~R.~A.} \bibnamefont{Palmans}},
  \bibnamefont{and} \bibinfo{author}{\bibfnamefont{E.~W.}
  \bibnamefont{Meijer}}, \bibinfo{journal}{Macromolecules}
  \textbf{\bibinfo{volume}{44}}, \bibinfo{pages}{6581–6587}
  (\bibinfo{year}{2011}).

\bibitem[{\citenamefont{Vedenov et~al.}(1972)\citenamefont{Vedenov, Dykhne, and
  D.}}]{DNA1}
\bibinfo{author}{\bibfnamefont{A.}~\bibnamefont{Vedenov}},
  \bibinfo{author}{\bibfnamefont{A.~M.} \bibnamefont{Dykhne}},
  \bibnamefont{and} \bibinfo{author}{\bibfnamefont{F.-K.~M.} \bibnamefont{D.}},
  \bibinfo{journal}{Sov. Phys. Usp.} \textbf{\bibinfo{volume}{14}},
  \bibinfo{pages}{715} (\bibinfo{year}{1972}).

\bibitem[{\citenamefont{Lifshitz}(1974)}]{DNA2}
\bibinfo{author}{\bibfnamefont{I.~M.} \bibnamefont{Lifshitz}},
  \bibinfo{journal}{Sov. Phys. JETP} \textbf{\bibinfo{volume}{38}},
  \bibinfo{pages}{545} (\bibinfo{year}{1974}).

\bibitem[{\citenamefont{Yashin et~al.}(1997)\citenamefont{Yashin, Kudryavtsev,
  Govorun, and Litmanovich}}]{Yashin}
\bibinfo{author}{\bibfnamefont{V.}~\bibnamefont{Yashin}},
  \bibinfo{author}{\bibfnamefont{Y.}~\bibnamefont{Kudryavtsev}},
  \bibinfo{author}{\bibfnamefont{E.}~\bibnamefont{Govorun}}, \bibnamefont{and}
  \bibinfo{author}{\bibfnamefont{A.}~\bibnamefont{Litmanovich}},
  \bibinfo{journal}{Macromol. Theory Simul.} \textbf{\bibinfo{volume}{6}},
  \bibinfo{pages}{247} (\bibinfo{year}{1997}).

\bibitem[{\citenamefont{Litmanovich et~al.}(2003)\citenamefont{Litmanovich,
  Kudryavtsev, Kriksin, and Kononenko}}]{Litmanovich1}
\bibinfo{author}{\bibfnamefont{A.~D.} \bibnamefont{Litmanovich}},
  \bibinfo{author}{\bibfnamefont{Y.~V.} \bibnamefont{Kudryavtsev}},
  \bibinfo{author}{\bibfnamefont{Y.~A.} \bibnamefont{Kriksin}},
  \bibnamefont{and} \bibinfo{author}{\bibfnamefont{O.~A.}
  \bibnamefont{Kononenko}}, \bibinfo{journal}{Macromol. Theory Simul.}
  \textbf{\bibinfo{volume}{12}}, \bibinfo{pages}{11} (\bibinfo{year}{2003}).

\bibitem[{\citenamefont{Litmanovich et~al.}(2010)\citenamefont{Litmanovich,
  Podbelskiy, and Kudryavtsev}}]{Litmanovich2}
\bibinfo{author}{\bibfnamefont{A.~D.} \bibnamefont{Litmanovich}},
  \bibinfo{author}{\bibfnamefont{V.~V.} \bibnamefont{Podbelskiy}},
  \bibnamefont{and} \bibinfo{author}{\bibfnamefont{Y.~V.}
  \bibnamefont{Kudryavtsev}}, \bibinfo{journal}{Macromol. Theory Simul.}
  \textbf{\bibinfo{volume}{19}}, \bibinfo{pages}{269} (\bibinfo{year}{2010}).

\bibitem[{\citenamefont{Smulders et~al.}(2010)\citenamefont{Smulders, Filot,
  van~der Schoot, Schenning, and Meijer}}]{Smulders1}
\bibinfo{author}{\bibfnamefont{M.~M.~J.} \bibnamefont{Smulders}},
  \bibinfo{author}{\bibfnamefont{J.~M.~A.} \bibnamefont{Filot},
  \bibfnamefont{I.~A. W.~Leenders}}, \bibinfo{author}{\bibfnamefont{P.~A.
  R.~A.} \bibnamefont{van~der Schoot}, \bibfnamefont{P}},
  \bibinfo{author}{\bibfnamefont{A.~P. H.~J.} \bibnamefont{Schenning}},
  \bibnamefont{and} \bibinfo{author}{\bibfnamefont{E.~W.}
  \bibnamefont{Meijer}}, \bibinfo{journal}{J. Am. Chem. Soc.}
  \textbf{\bibinfo{volume}{132}}, \bibinfo{pages}{611–619}
  (\bibinfo{year}{2010}).

\bibitem[{\citenamefont{Nyrkova et~al.}(2000)\citenamefont{Nyrkova, Semenov,
  Aggeli, Bell, Boden, and McLeish}}]{nyrkova}
\bibinfo{author}{\bibfnamefont{I.}~\bibnamefont{Nyrkova}},
  \bibinfo{author}{\bibfnamefont{A.}~\bibnamefont{Semenov}},
  \bibinfo{author}{\bibfnamefont{A.}~\bibnamefont{Aggeli}},
  \bibinfo{author}{\bibfnamefont{M.}~\bibnamefont{Bell}},
  \bibinfo{author}{\bibfnamefont{N.}~\bibnamefont{Boden}}, \bibnamefont{and}
  \bibinfo{author}{\bibfnamefont{T.}~\bibnamefont{McLeish}},
  \bibinfo{journal}{Eur. Phys. J. B} \textbf{\bibinfo{volume}{17}},
  \bibinfo{pages}{499} (\bibinfo{year}{2000}).

\bibitem[{\citenamefont{van~der Schoot}(2005)}]{SupraSA2}
\bibinfo{author}{\bibfnamefont{P.}~\bibnamefont{van~der Schoot}},
  \emph{\bibinfo{title}{Supramolecular Polymers}} (\bibinfo{publisher}{CRC
  Press}, \bibinfo{year}{2005}), chap. \bibinfo{chapter}{2. Theory of
  Supramolecular Polymerization}, pp. \bibinfo{pages}{77–--106},
  \bibinfo{edition}{2nd} ed.

\bibitem[{\citenamefont{van~der Schoot}(2009)}]{Paul}
\bibinfo{author}{\bibfnamefont{P.}~\bibnamefont{van~der Schoot}},
  \emph{\bibinfo{title}{Advances In Chemical Engineering.}}
  (\bibinfo{publisher}{Academic Press}, \bibinfo{year}{2009}),
  vol.~\bibinfo{volume}{35}, chap. \bibinfo{chapter}{3. Nucleation and
  Co-Operativity in Supramolecular Polymers}, pp. \bibinfo{pages}{45--77}.

\bibitem[{\citenamefont{Zimm and Bragg}(1957)}]{Zimm-Bragg}
\bibinfo{author}{\bibfnamefont{B.~H.} \bibnamefont{Zimm}} \bibnamefont{and}
  \bibinfo{author}{\bibfnamefont{J.~K.} \bibnamefont{Bragg}},
  \bibinfo{journal}{{J. Chem. Phys.}} \textbf{\bibinfo{volume}{31}},
  \bibinfo{pages}{526} (\bibinfo{year}{1957}).

\bibitem[{\citenamefont{Rubinstein and Colby}(2003)}]{Rubinstein}
\bibinfo{author}{\bibfnamefont{M.}~\bibnamefont{Rubinstein}} \bibnamefont{and}
  \bibinfo{author}{\bibfnamefont{R.}~\bibnamefont{Colby}},
  \emph{\bibinfo{title}{Polymer Physics}} (\bibinfo{publisher}{Oxford
  University Press}, \bibinfo{year}{2003}).

\bibitem[{\citenamefont{Vermonden et~al.}(2003)\citenamefont{Vermonden, van~der
  Gucht, de~Waard, Marcelis, Besseling, Fleer, and Cohen~Stuart}}]{Vermonden}
\bibinfo{author}{\bibfnamefont{T.}~\bibnamefont{Vermonden}},
  \bibinfo{author}{\bibfnamefont{J.}~\bibnamefont{van~der Gucht}},
  \bibinfo{author}{\bibfnamefont{P.}~\bibnamefont{de~Waard}},
  \bibinfo{author}{\bibfnamefont{A.}~\bibnamefont{Marcelis}},
  \bibinfo{author}{\bibfnamefont{E.}~\bibnamefont{Besseling},
  \bibfnamefont{N.A.M.~Sudhölter}},
  \bibinfo{author}{\bibfnamefont{G.}~\bibnamefont{Fleer}}, \bibnamefont{and}
  \bibinfo{author}{\bibfnamefont{M.}~\bibnamefont{Cohen~Stuart}},
  \bibinfo{journal}{Macromolecules} \textbf{\bibinfo{volume}{36}},
  \bibinfo{pages}{7035} (\bibinfo{year}{2003}).

\bibitem[{\citenamefont{Goldenfeld}(1992)}]{goldenfeld}
\bibinfo{author}{\bibfnamefont{N.}~\bibnamefont{Goldenfeld}},
  \emph{\bibinfo{title}{Lectures on Phase Transitions and Renormalization
  Group}} (\bibinfo{publisher}{Adison Wesley Publishing Company},
  \bibinfo{year}{1992}), \bibinfo{edition}{1st} ed.

\bibitem[{\citenamefont{Flory}(1953)}]{Flory}
\bibinfo{author}{\bibfnamefont{P.~J.} \bibnamefont{Flory}},
  \emph{\bibinfo{title}{Principles of Polymer Chemistry}}
  (\bibinfo{publisher}{Cornell University Press: New York},
  \bibinfo{year}{1953}).

\bibitem[{\citenamefont{Sollich}(2002)}]{Sollich}
\bibinfo{author}{\bibfnamefont{P.}~\bibnamefont{Sollich}},
  \bibinfo{journal}{Journal of Physics: Condensed Matter}
  \textbf{\bibinfo{volume}{14}}, \bibinfo{pages}{R79} (\bibinfo{year}{2002}).

\bibitem[{\citenamefont{Wilding et~al.}(2008)\citenamefont{Wilding, Sollich,
  and Buzzacchi}}]{Sollich1}
\bibinfo{author}{\bibfnamefont{N.~B.} \bibnamefont{Wilding}},
  \bibinfo{author}{\bibfnamefont{P.}~\bibnamefont{Sollich}}, \bibnamefont{and}
  \bibinfo{author}{\bibfnamefont{M.}~\bibnamefont{Buzzacchi}},
  \bibinfo{journal}{Phys. Rev. E} \textbf{\bibinfo{volume}{77}},
  \bibinfo{pages}{011501} (\bibinfo{year}{2008}).

\bibitem[{\citenamefont{Jabbari-Farouji and Sollich}()}]{Sollich2}
\bibinfo{author}{\bibfnamefont{S.}~\bibnamefont{Jabbari-Farouji}}
  \bibnamefont{and} \bibinfo{author}{\bibfnamefont{P.}~\bibnamefont{Sollich}},
  \bibinfo{note}{in preparation}.

\end{thebibliography}

\end {document}